\documentclass[english,aps,pra,preprint,showpacs]{revtex4-2}
\usepackage[T1]{fontenc}
\usepackage[latin9]{inputenc}
\usepackage{color}
\usepackage{babel}
\usepackage{float}
\usepackage{amsmath}
\usepackage{graphicx}
\usepackage[pdfusetitle,
 bookmarks=true,bookmarksnumbered=false,bookmarksopen=false,
 breaklinks=false,pdfborder={0 0 1},backref=false,colorlinks=true]
 {hyperref}
\hypersetup{
 pdfborderstyle=}

\makeatletter
\usepackage{babel}

\makeatother

\begin{document}
\title{Multipartite continuous-variable quantum nondemolition interaction
and entanglement certification and monitoring}
\author{Vinícius V. Seco}
\author{Alencar J. de Faria}
\email{alencar.faria@unifal-mg.edu.br}

\affiliation{Instituto de Ciência e Tecnologia, Universidade Federal de Alfenas,
CEP 37715-400, Poços de Caldas, MG, Brazil}
\begin{abstract}
The quantum nondemolition (QND) measurement is one of the most studied
quantum measurement procedures. Usually, such a process involves the
coupling of a single system of interest, called signal, with a single
probe system, so that the relevant information in the signal system
is indirectly measured by observing the probe system. Here, we extend
the concept of quantum nondemolition interaction to the cases in which
the signal and the probe systems are each one multipartite continuous-variable
systems. Specifically, we propose a general scheme that performs the
multipartite QND interactions, relying on beam-splitter couplings
among the signal and probe modes with only two ancillary modes prepared
off-line in squeezed states. The scheme is also composed of homodyne
detections and feedforward modulations. The ancillary modes are detected
in the process, providing photocurrents for post-modulation of the
output systems, as well as sufficient information to calculate genuine
multipartite entanglement conditions of the input systems and to monitor
similar conditions of the output systems. 
\end{abstract}
\maketitle

\section{Introduction}

A highly relevant and useful physical process in Quantum Information
is to be able to perform an indirect measurement of a given variable
of a system of interest without specifically altering the dynamical
evolution of this same variable. For this task there is a type of
procedure called quantum nondemolition (QND) measurement \citep{Braginsky80,Braginsky96,Grangier98,Wiseman-Milburn}.
In this process, a probe system interacts with a signal system in
order to produce a correlated state between them. The overall Hamiltonian
follows some conditions, so that the evolution of a signal observable
is not perturbed, but a probe observable is changed, receiving information
just from the unperturbed signal observable. Then the probe system
can be measured in a suitable state base for obtaining information
about the signal system. This procedure is very relevant for repeatedly
obtaining high-precision measurements of a given variable, to the
detriment of the information loss of the respective conjugate variable.
Applications of QND measurements have recently been found for detections
of single photon or electron spin in many different systems \citep{Besse18,Vasilyev20,Yoneda20},
and in measurement and manipulation of collective atomic spins \citep{Eckert07,ReviewMa11,Yang20}.
In the scenario of continuous-variable systems \citep{Braunstein05},
the QND measurement and squeezed laser beams have been implemented
in more recent apparatus of gravitational wave detectors \citep{Schnabel17,LIGO23}.

Early experimental implementations of the QND measurements were made
by strongly pumped nonlinear media, as optical crystals in cavities
or optical fibers, where the probe and signal modes can be coupled
\citep{Milburn83,Imoto85,Levenson86,Yurke85,La=000020Porta89,Scully}.
Another method relying on the application of feedforward loops on
the modes. Coupling the modes of interest with an ancillary mode prepared
off-line in a squeezed state and so detecting the ancillary mode by
homodyne measurement, the resulting photocurrent can activate modulations
on the modes of interest, closing a feedforward loop \citep{Lam97,Ralph97}.
This strategy was developed and experimentally demonstrated for bipartite
QND interaction \citep{Andersen02,Filip05,Yoshikawa08,Marek10,Shiozawa18},
mode squeezing \citep{Buchler01,Yoshikawa07,Miyata14}, quantum computation
and processing \citep{Ukai11,Yokoyama15,Takeda19,Sakaguchi23}, production
of nonlinear interaction \citep{Marek11,Miyata16,Marek18,Sefi19},
and nondestructive certification of bipartite entanglement \citep{Faria16}.
The multipartite QND interaction scheme proposed in the present article
is based on these measurement-mediated and feedforward loop strategies.

Since a QND measurement requires an interaction that produces correlated
states between the probe and signal systems, it is natural to be concerned
with monitoring the conditions of entanglement or other quantum correlation
before and after the QND interaction. A great challenge in current
quantum information research is how to detect, quantify and manipulate
quantum correlations of the multipartite systems. Among the most studied
quantum correlations, we can cite the entanglement \citep{Horodecki09}
and the EPR steering \citep{Uola20}. Recent efforts have been accomplished
to characterize multipartite entanglement and EPR steering in the
context of continuous-variable systems \citep{Giedke01,Loock03,Sperling13,He13,Teh14,Shchukin15,Teh22}.
Most of those studies relied on the statistical properties of the
systems, especially the variances of dynamical operators, to obtain
sufficient conditions for identifying of the correlations in quantum
regime\citep{Loock03,He13,Teh14,Shchukin15,Teh22}.

Criteria for detecting entanglement of continuous-variable modes have
been widely established for bipartite systems, with necessary and
sufficient conditions for Gaussian systems, but only sufficient for
non-Gaussian cases \citep{Simon00,DGCZ00,Giovannetti03}. For multipartite
systems, there are major complications regarding the entanglement
structure and their depth, that is, how many and which parts are entangled
with each other. A particularly interesting case of entanglement is
the so-called genuine multipartite entanglement, in which the state
of the complete system can not be described as a statistical mixture
of biseparable states, that is, those formed by the tensor product
of the states of two sets of subsystems. In this article we will focus
on studying such genuine entanglement states in the context of continuous-variable
systems, for which there are many works in the literature establishing
sufficient detection criteria. Specifically, we will follow the works
that use the definition of genuine entanglement mentioned above \citep{Teh14}.
Associated with these studies of genuine quantum correlation, one
must mention criteria for genuine multipartite EPR steering, which
would fit into the study to be presented in this article, but will
be left for a later work.

In this article, we will extend the concept of QND interaction between
two systems, called the signal and the probe ones, to the case where
both systems are sets of multipartite subsystems. A multipartite QND
interaction must alter many probe systems, projecting onto them the
states from many signal systems, while preserving signal observables
of interest. Few works have studied multipartite QND processes in
recent literature \citep{Yang20,Sefi19}, in the sense of dealing
with a single simultaneous operation on independent systems, as opposed
to treating collective systems as single one or many operations carried
out in parallel. Thus, we propose a scheme formulated for modes in
the continuous variable context, so that a QND interaction is performed
among many multipartite systems. For its implementation, two ancillary
modes previously prepared off-line in squeezed states are coupled
to the signal and probe systems by a set of beam splitters. After
both ancillary modes couple with all the multipartite modes, they
are separately measured in conjugate quadratures by homodyne detections.
Both photocurrents generated by detectors are used for feeding subsequent
modulations on the signal and probe modes, closing feedforward loops
and accomplishing the multipartite QND interaction. There is a second
purpose for the detected photocurrents, that is to use them to calculate
covariance values between the multipartite signal and probe modes.
With these data, it is possible to sufficiently certify the genuine
entanglement of the input or output modes. The criteria calculated
for multipartite genuine entanglement certification \citep{Teh14}
are sufficient conditions to both Gaussian and non Gaussian modes.
In what follows, we will call the set of signal and probe modes as
target modes, in opposition to the ancillary modes. 

\section{Multipartite Quantum Nondemolition Interaction }

\subsection{General case}

A generalization for quantum nondemolition (QND) interactions is to
consider both signal and probe systems as composed of multipartite
modes. There are many different realizations of QND interactions between
one signal and one probe systems. Here, we are based on a parametric
coupling between modes described by conjugated quadrature operators
$\hat{q}_{S}$ and $\hat{p}_{S}$, for the signal system, and $\hat{q}_{P}$
and $\hat{p}_{P}$, for the probe system, such that
\begin{eqnarray}
\hat{q}_{S}^{\mathrm{(out)}} & = & \hat{q}_{S}^{\mathrm{(in)}}\label{QND-qs}\\
\hat{p}_{S}^{\mathrm{(out)}} & = & \hat{p}_{S}^{\mathrm{(in)}}-G\hat{p}_{P}^{\mathrm{(in)}}\label{QND-ps}\\
\hat{q}_{P}^{\mathrm{(out)}} & = & \hat{q}_{P}^{\mathrm{(in)}}+G\hat{q}_{S}^{\mathrm{(in)}}\label{QND-qp}\\
\hat{p}_{P}^{\mathrm{(out)}} & = & \hat{p}_{P}^{\mathrm{(in)}},\label{QND-pp}
\end{eqnarray}
 where the indexes $\mathrm{(in)}$ and $\mathrm{(out)}$ mean the
operator before and after the interaction, respectively, and $G$
is a gain of the interaction \citep{Grangier98,Imoto85,Levenson86,Yurke85,La=000020Porta89,Scully}.
The quadrature operators can be written in terms of annihilation and
creation operators, $\hat{a}_{j}$ and $\hat{a}_{j}^{\dagger}$, with
$j=S;P$, and initial phases $\theta_{j}$, such that $\hat{q}_{j}^{\mathrm{(in)}}=\frac{1}{\sqrt{2}}(e^{-i\theta_{j}}\hat{a}_{j}+e^{i\theta_{j}}\hat{a}_{j}^{\dagger})$
and $\hat{p}_{j}^{\mathrm{(in)}}=\frac{1}{i\sqrt{2}}(e^{-i\theta_{j}}\hat{a}_{j}-e^{i\theta_{j}}\hat{a}_{j}^{\dagger})$,
obeying the commutation relations $[\hat{a}_{j},\hat{a}_{k}^{\dagger}]=\delta_{jk}$
and $[\hat{a}_{j},\hat{a}_{k}]=[\hat{a}_{j}^{\dagger},\hat{a}_{k}^{\dagger}]=0$. 

We can obtain similar equations to Eqs. \eqref{QND-qs}--\eqref{QND-pp}
in the case of multipartite signal and probe systems. To construct
such an equation set, consider a complete continuous-variable system
compounded by $N$ modes, so that each mode is represented by a pair
of conjugated quadrature operators. For the j-th mode, we arrange
the operators as 
\begin{equation}
\hat{X}_{j}^{\mathrm{(in)}}=\left(\begin{array}{c}
\hat{q}_{j}^{\mathrm{(in)}}\\
\hat{p}_{j}^{\mathrm{(in)}}
\end{array}\right).\label{mode-in}
\end{equation}
 The coupling Hamiltonian to produce a QND interaction among $N$
modes must be
\begin{equation}
\hat{H}_{I}=\sum_{j=1}^{m}\sum_{k=m+1}^{N}\hbar c_{jk}\hat{q}_{j}^{\mathrm{(in)}}\hat{p}_{k}^{\mathrm{(in)}},\label{hamiltonian-qnd}
\end{equation}
so that the first $m$ modes are attributed to the signal modes and
following $N-m$ modes are attributed to the probe modes. The dynamical
evolution of the Hamiltonian \eqref{hamiltonian-qnd} corresponds
to a unitary operator 
\begin{equation}
\hat{U}=\exp\left(-i\sum_{j=1}^{m}\sum_{k=m+1}^{N}G_{jk}\hat{q}_{j}^{\mathrm{(in)}}\hat{p}_{k}^{\mathrm{(in)}}\right).\label{unitary-qnd}
\end{equation}
where $G_{jk}=c_{jk}\tau$ and $\tau$ is the coupling period. This
dynamical evolution application applied to input quadrature operators
returns the respective output operators, $\hat{X}_{j}^{\mathrm{(out)}}=\hat{U}^{\dagger}\hat{X}_{j}^{\mathrm{(in)}}\hat{U}$.
Fortunately the operators of different modes in Eq. \eqref{unitary-qnd}
commute with each other, so
\begin{equation}
\hat{U}=\prod_{j=1}^{m}\prod_{k=m+1}^{N}\exp\left(-iG_{jk}\hat{q}_{j}^{\mathrm{(in)}}\hat{p}_{k}^{\mathrm{(in)}}\right).\label{unitary-qnd-prod}
\end{equation}
Proceeding with the usual calculations, we obtain for the signal modes
\begin{equation}
\hat{X}_{j}^{\mathrm{(out)}}=\left(\begin{array}{c}
\hat{q}_{j}^{\mathrm{(out)}}\\
\hat{p}_{j}^{\mathrm{(out)}}
\end{array}\right)=\left(\begin{array}{c}
\hat{q}_{j}^{\mathrm{(in)}}\\
\hat{p}_{j}^{\mathrm{(in)}}-\sum_{k=m+1}^{N}G_{jk}\hat{p}_{k}^{\mathrm{(in)}}
\end{array}\right),\label{signalmode}
\end{equation}
to mode $j$ with $1\leq j\leq m$; and for the probe modes
\begin{equation}
\hat{X}_{k}^{\mathrm{(out)}}=\left(\begin{array}{c}
\hat{q}_{k}^{\mathrm{(out)}}\\
\hat{p}_{k}^{\mathrm{(out)}}
\end{array}\right)=\left(\begin{array}{c}
\hat{q}_{k}^{\mathrm{(in)}}+\sum_{j=1}^{m}G_{jk}\hat{q}_{j}^{\mathrm{(in)}}\\
\hat{p}_{k}^{\mathrm{(in)}}
\end{array}\right)\label{probemode}
\end{equation}
to mode $k$ with $m+1\leq k\leq N$. This outcome shows $m$ modes
with their statistics unchanged in quadratures $\hat{q}_{j}^{\mathrm{(in)}}$
and carrying information from the other modes in their transformed
quadratures $\hat{p}_{j}^{\mathrm{(in)}}$. In the other hand, the
opposite occurs to the other $N-m$ modes, but with exchanged quadratures. 

\subsection{QND interaction with a single probe mode}

Let us study a particular case, which $N-1$ signal modes interact
with a single probe mode, so that the QND interaction yields in $N-1$
mode quadratures with unchanged statistics, at the cost of stain the
respective conjugated quadratures with the noise provided from the
probe mode. On the other hand, after the QND interaction, a single
probe mode quadrature carries the information from all other modes.
We will present a QND interaction restricted to a full continuous-variable
system, that alludes to a theoretical and experimental work of Yang
et al. \citep{Yang20}. In that work, it has been implemented a QND
measurement to a trapped-ion simulator made up of a multipartite spin
system, coupled with their center-of-mass vibrational mode, playing
as the probe system. Thus such a setting performs a hybrid QND interaction
between a multipartite discrete variable signal system and a single
continuous-variable mode. 

Returning to the case of full continuous-variable QND interaction,
the Hamiltonian for multipartite signal modes and a single probe mode
must be, from Eq. \eqref{hamiltonian-qnd}, 
\begin{equation}
\hat{H}_{I}=\sum_{j=1}^{N-1}\hbar c_{j}\hat{q}_{j}^{\mathrm{(in)}}\hat{p}_{N}^{\mathrm{(in)}},\label{hamiltonian-qnd-1probe}
\end{equation}
so that the respective unitary mapping, after the interaction period,
is 
\begin{equation}
\hat{U}=\exp\left(-i\sum_{j=1}^{N-1}G_{j}\hat{q}_{j}^{\mathrm{(in)}}\hat{p}_{N}^{\mathrm{(in)}}\right)=\prod_{j=1}^{N-1}\exp\left(-iG_{j}\hat{q}_{j}^{\mathrm{(in)}}\hat{p}_{N}^{\mathrm{(in)}}\right).\label{unitary-qnd-1probe}
\end{equation}
Hence, the output modes can be calculated in terms of the input modes,
such that 
\begin{equation}
\hat{X}_{j}^{\mathrm{(out)}}=\left(\begin{array}{c}
\hat{q}_{j}^{\mathrm{(out)}}\\
\hat{p}_{j}^{\mathrm{(out)}}
\end{array}\right)=\left(\begin{array}{c}
\hat{q}_{j}^{\mathrm{(in)}}\\
\hat{p}_{j}^{\mathrm{(in)}}-G_{j}\hat{p}_{N}^{\mathrm{(in)}}
\end{array}\right),\label{signalmode-1probe}
\end{equation}
to the j-th signal mode, with $1\leq j<N-1$. The probe mode, labeled
with index $N$, results in
\begin{equation}
\hat{X}_{N}^{\mathrm{(out)}}=\left(\begin{array}{c}
\hat{q}_{N}^{\mathrm{(out)}}\\
\hat{p}_{N}^{\mathrm{(out)}}
\end{array}\right)=\left(\begin{array}{c}
\hat{q}_{N}^{\mathrm{(in)}}+\sum_{j=1}^{N-1}G_{j}\hat{q}_{j}^{\mathrm{(in)}}\\
\hat{p}_{N}^{\mathrm{(in)}}
\end{array}\right).\label{probemode-1probe}
\end{equation}

These N-mode operations are a multipartite generalization of QND interaction
with a single probe, but they can also be understood as a related
case of multipartite nonlinear coupling studied in references \citep{Sefi19,Hanamura24}.
Moreover, in those proposals, the N-mode multipartite interaction
is performed with the use of $N$ ancillary modes, which are off-line
prepared in non-Gaussian states. In the next section, we present the
feasibility of coupling $N$ target modes as in Eqs. \eqref{signalmode-1probe}
and \eqref{probemode-1probe}, using only 2 ancillary modes prepared
in Gaussian squeezed states.

\section{N-partite QND Interaction Scheme}

For the implementation of multipartite continuous-variable QND interaction,
we consider a general scheme based on the measurement-mediated method
from previous works \citep{Filip05,Yoshikawa08,Marek10,Shiozawa18,Sefi19,Faria16,Hanamura24}.
The proposed scheme is flexible, so it can have many variants intended
for different purposes. Here we show the scheme design and detail
a few remarkable cases. Let us consider a system composed of $N$
target modes, indexed from $1$ to $N$, which $n$ modes are considered
signal ones and the other $N-n$ modes are the probe ones. Unlike
applying the QND interaction to each pair of signal and probe modes,
what would be very expensive, we consider only two ancillary modes
prepared in known squeezed states. The ancillary modes, labeled with
indexes A and B, can be two independent squeezed states or entangled
two-mode squeezed states, and must be coupled to each target mode
by photon-conserving linear operations, as beam splitters. After that,
the conjugate quadratures from ancillary modes are measured by homodyne
detection, so that the photoelectric signals are used in a feedforward
modulation loop on the target modes, concluding the multipartite QND
interaction.

As the detected photoelectric signals have classical character, we
get an extra resource in the procedure. Their statistical properties
can be analyzed to obtain variances of the ancillary mode quadratures,
then obtaining variances of the linear combinations of the target
mode quadratures. So, the measurement outcomes can be used to certificate
quantum correlations of the target modes, such as genuine multipartite
entanglement, as the conditions obtained by references \citep{Teh14,Shchukin15,Armstrong},
or EPR steering in references \citep{He13,Teh22,Armstrong}. 

\begin{figure}[H]
\begin{centering}
\includegraphics[width=16cm]{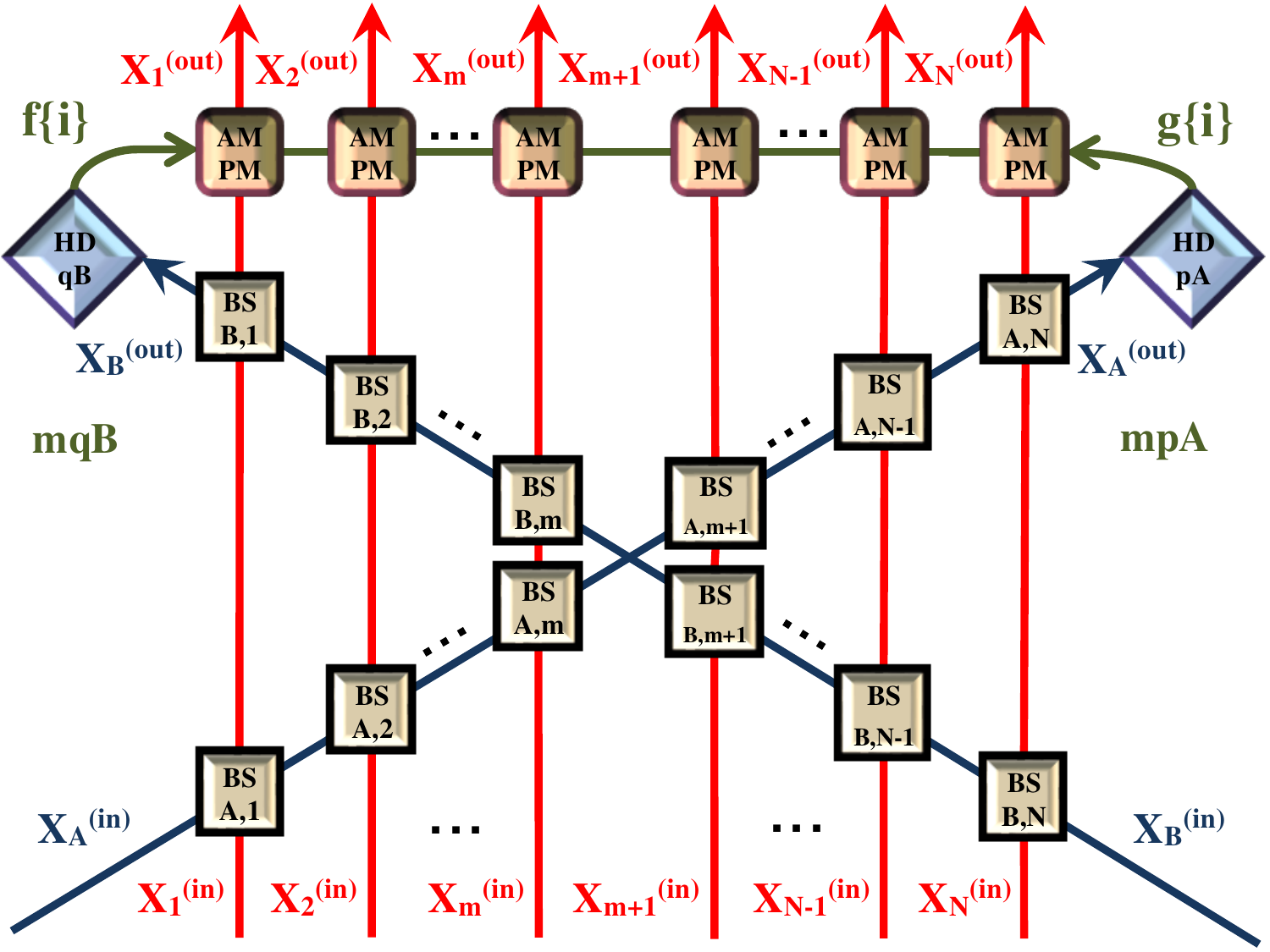}
\par\end{centering}
\centering{}\caption{Scheme for multipartite QND interaction among target modes labeled
from $X_{1}^{\mathrm{(in)}}$ to $X_{N}^{\mathrm{(in)}}$. The process
is performed coupling all target modes to two ancillary modes, $X_{A}^{\mathrm{(in)}}$
and $X_{B}^{\mathrm{(in)}}$, by beam-splitter operations labeled
with BS and indexes from the respective modes. After the all beam-splitter
operations, the quadratures operators $\hat{p}_{A}^{\mathrm{(out)}}$
and $\hat{q}_{B}^{\mathrm{(out)}}$ from ancillary modes are measured
by homodyne detections, HD, whose outcomes provide control to modulate
the target modes. The AM PM modulations add the detected signals from
$\hat{p}_{A}^{\mathrm{(out)}}$ and $\hat{q}_{B}^{\mathrm{(out)}}$
to j-th target mode with gains $g_{j}$ and $f_{j}$ respectively.
So the output target modes $X_{\{j\}}^{\mathrm{(out)}}$ are obtained,
reproducing a multipartite QND interaction. \protect\label{scheme}}
\end{figure}

As shown in Figure \ref{scheme}, the QND interaction implementation
is made by coupling the $N$ target modes with two ancillary modes
A and B using beam-splitter operations. There are many ways to arrange
the couplings between mode pairs, but the most direct is crossing
the ancillary modes along the target beams from opposite sides \citep{Faria16}. 

The beam-splitter operations are featured by transmission and reflection
coefficients, $t$ and $r$, related by $t^{2}+r^{2}=1$ \citep{Scully}.
Each one is labeled with 2 indexes, designating the respective coupled
mode pair, A or B from the ancillary modes and $1\leq j\leq N$ from
the target modes. 

The ancillary mode $A$ has the initial quadrature operators denoted
by $\hat{X}_{A}^{\mathrm{(in)}}$. After each coupling with the j-th
target mode with quadratures $\hat{X}_{j}^{\mathrm{(in)}}$, the resulting
ancillary mode is $\hat{X}_{A}^{(j)}$ and the i-th target mode is
$\hat{X}_{j}^{(1)}$, so that the beam-splitter coupling results in
\begin{eqnarray}
\hat{X}_{j}^{(1)} & = & t_{A,j}\hat{X}_{j}^{\mathrm{(in)}}+r_{A,j}\hat{X}_{A}^{(j-1)}\label{xita}\\
\hat{X}_{A}^{(j)} & = & t_{A,j}\hat{X}_{A}^{(j-1)}-r_{A,j}\hat{X}_{j}^{\mathrm{(in)}}\label{xata}
\end{eqnarray}
for $1\leq j\leq m$. In the other side, beginning from the N-th target
mode, the ancillary mode $B$, with initial quadrature operators $\hat{X}_{B}^{\mathrm{(in)}}$,
couples with the j-th target mode, yielding the ancillary and j-th
target quadratures $\hat{X}_{B}^{(j)}$ and $\hat{X}_{j}^{(1)}$,
respectively, so that
\begin{eqnarray}
\hat{X}_{j}^{(1)} & = & t_{B,j}\hat{X}_{j}^{\mathrm{(in)}}+r_{B,j}\hat{X}_{B}^{(N-j)}\label{xjtb}\\
\hat{X}_{B}^{(N+1-j)} & = & t_{B,j}\hat{X}_{B}^{(N-j)}-r_{B,j}\hat{X}_{j}^{\mathrm{(in)}}\label{xbtb}
\end{eqnarray}
for $m+1\leq j\leq N$. These beam-splitter coupling perform the first
information gathering by ancillary modes.

After that, the ancillary modes A and B exchange the target modes
with which perform the second round of beam-splitter couplings. So
the resulting modes are
\begin{eqnarray}
\hat{X}_{j}^{(2)} & = & t_{B,j}\hat{X}_{j}^{(1)}+r_{B,j}\hat{X}_{B}^{(N-j)}\label{xitb2}\\
\hat{X}_{B}^{(N+1-j)} & = & t_{B,j}\hat{X}_{B}^{(N-j)}-r_{B,j}\hat{X}_{j}^{(1)}\label{xbtb2}
\end{eqnarray}
for $1\leq j\leq m$, and
\begin{eqnarray}
\hat{X}_{j}^{(2)} & = & t_{A,j}\hat{X}_{j}^{(1)}+r_{A,j}\hat{X}_{A}^{(j-1)}\label{xjta2}\\
\hat{X}_{A}^{(j)} & = & t_{A,j}\hat{X}_{A}^{(j-1)}-r_{A,j}\hat{X}_{j}^{(1)}\label{xata2}
\end{eqnarray}
for $m+1\leq j\leq N$. 

This procedure gives us two ancillary modes that interacted with all
target modes. With both ancillary modes, one pair of measurements
on conjugate quadratures can be made, using homodyne detections, e.g.,
on the quadrature $\hat{p}_{A}^{\mathrm{(out)}}$ from the output
ancillary mode $\hat{X}_{A}^{(N)}=\hat{X}_{A}^{\mathrm{(out)}}$ and
on the quadrature $\hat{q}_{B}^{\mathrm{(out)}}$ from the output
ancillary mode $\hat{X}_{B}^{(N)}=\hat{X}_{B}^{\mathrm{(out)}}$.
The homodyne measurements generate photocurrents $I_{A}$ and $I_{B}$
proportional to the measured values $p_{A}^{\mathrm{(out)}}$ and
$q_{B}^{\mathrm{(out)}}$, respectively related to observables $\hat{p}_{A}^{\mathrm{(out)}}$
and $\hat{q}_{B}^{\mathrm{(out)}}$. These photocurrents feed post-modulations
on the target modes, adding their fluctuations to the quadratures
of each mode $j$ like $\hat{q}_{j}^{(2)}+\phi_{j}I_{B}$ and $\hat{p}_{j}^{(2)}+\gamma_{j}I_{A}$,
where $\phi_{j}$ and $\gamma_{j}$ are adjustable gains. Such random
modulations can be described and substituted by the respective quadrature
operators, for the purpose of calculating the output target states
\citep{Filip05,Yoshikawa08,Marek10,Shiozawa18}. To make the analysis
of the scheme clearer, we do not take into account the losses due
to the mode propagations and the detector efficiencies. Every noise
addition to output modes comes from the intrinsic quantum measurement
processes.

A fundamental aspect that must be taken into account is the necessity
of performing compatible measurements of the output ancillary modes,
that is, we must look for commutative observables. A reason is the
measurement of both output ancillary quadratures $\hat{p}_{A}^{\mathrm{(out)}}$
and $\hat{q}_{B}^{\mathrm{(out)}}$ can be obtained without uncertainty
inferior limit, reducing the measurement noise addition to ideally
zero \citep{Wiseman-Milburn,Filip05}. Another aspect is that the
photocurrents must carry the information from the same multipartite
state of the target modes, in order that they can feed post-modulations
on the conjugate target quadratures, producing output target modes
which are expressible by the same input operators. Otherwise, the
resulting target modes should be expressed by time-domain operators
and an analysis of modes with bandwidth would be necessary. Moreover,
the same photocurrents must provide data to calculate correlations
between the target modes, which must represent a single observed multipartite
state \citep{Loock03,He13,Teh14,Teh22,Simon00,DGCZ00,Giovannetti03},
as we will show in Section IV. These conditions will be placed on
the proposed scheme in what follows.

To demonstrate the viability of the proposed scheme, let us consider
the case of QND interactions with a single probe mode, as described
in Subsection II.B. This particular case can be performed choosing
suitable transmission and reflection coefficients. A first choice
would be to take all $t_{A,j}$ and $t_{B,j}$ equal (so $r_{A,j}$
and $r_{B,j}$ are equal too), but such situation results in incompatible
measurements of ancillary modes. Thus, at least one beam-splitter
operation must have a different transmission coefficient from the
others, and another one or the same operation must be out of phase
with respect to others. So a simple choice is to take 
\begin{equation}
t_{A,i}=t_{B,j}=t_{o}\quad\mbox{, for \ensuremath{i,j\neq N}},\label{transmission-o}
\end{equation}
\begin{equation}
t_{A,N}=t_{B,N}=t_{d},\label{transmission-d}
\end{equation}
 so that the reflection coefficients are as usual $r_{A,i}=r_{B,j}=r_{o}=\sqrt{1-t_{o}^{2}}$,
for any $i,j\neq N$, but in the N-mode line both reflection coefficients
must be
\begin{equation}
r_{B,N}=-r_{A,N}=r_{d}=\sqrt{1-t_{d}^{2}}.\label{reflection-d}
\end{equation}

With the setup determined by Eqs. \eqref{transmission-o}--\eqref{reflection-d},
the ancillary modes A and B are measured by homodyne detections, yielding
measured quadratures $\hat{p}_{A}^{\mathrm{(out)}}$ and $\hat{q}_{B}^{\mathrm{(out)}}$
given by
\begin{equation}
\hat{p}_{A}^{\mathrm{(out)}}=t_{d}t_{o}^{N-1}\left(\sum_{j=1}^{N-1}f_{j}\hat{p}_{j}^{\mathrm{(in)}}-f_{N}\hat{p}_{N}^{\mathrm{(in)}}\right)+t_{d}t_{o}^{N-1}\hat{p}_{A}^{\mathrm{(in)}}+\left\{ 1-t_{d}^{2}\left[2-t_{o}^{2(N-m-1)}\right]\right\} \hat{p}_{B}^{\mathrm{(in)}}\label{pain}
\end{equation}
and 
\begin{equation}
\hat{q}_{B}^{\mathrm{(out)}}=t_{d}t_{o}^{N-1}\left(\sum_{j=1}^{N-1}g_{j}\hat{q}_{j}^{\mathrm{(in)}}-g_{N}\hat{q}_{N}^{\mathrm{(in)}}\right)-(1-t_{o}^{2m})\hat{q}_{A}^{\mathrm{(in)}}+t_{d}t_{o}^{N-1}\hat{q}_{B}^{\mathrm{(in)}},\label{qbin}
\end{equation}
where the following coefficients are defined,
\begin{eqnarray}
f_{j} & = & -r_{o}t_{o}^{-j},\label{gi}\\
g_{j} & = & -r_{o}t_{d}^{-1}t_{o}^{2m-N-j+1},\label{fi}
\end{eqnarray}
 for $1\leq j\leq m$, 
\begin{eqnarray}
f_{j} & = & -r_{o}t_{o}^{j-2m-1},\label{gj}\\
g_{j} & = & -r_{o}t_{d}^{-1}t_{o}^{j-N},\label{fj}
\end{eqnarray}
 for $m+1\leq j\leq N-1$, and 
\begin{eqnarray}
f_{N} & = & -r_{d}t_{o}^{1-N}\left[2-t_{o}^{2(N-m-1)}\right],\label{gN}\\
g_{N} & = & r_{d}t_{d}^{-1}.\label{fN}
\end{eqnarray}

On the other hand, after all beam-splitter interactions, the target
modes are found as
\begin{eqnarray}
\hat{X}_{j}^{(2)} & = & \hat{X}_{j}^{\mathrm{(in)}}-r_{o}^{2}\sum_{i=1}^{m}t_{o}^{2m-j-i}\hat{X}_{i}^{\mathrm{(in)}}-r_{o}^{2}\sum_{i=m+1}^{N-1}t_{o}^{-j-1+i}\hat{X}_{i}^{\mathrm{(in)}}-r_{o}r_{d}t_{o}^{N-j-1}\hat{X}_{N}^{\mathrm{(in)}}+\nonumber \\
 & + & r_{o}t_{o}^{2m-j}\hat{X}_{A}^{\mathrm{(in)}}+r_{o}t_{d}t_{o}^{N-j-1}\hat{X}_{B}^{\mathrm{(in)}}\label{xi2}
\end{eqnarray}
for $1\leq j\leq m$, 
\begin{eqnarray}
\hat{X}_{j}^{(2)} & = & \hat{X}_{j}^{\mathrm{(in)}}-r_{o}^{2}\sum_{i=1}^{m}t_{o}^{j-1-i}\hat{X}_{i}^{\mathrm{(in)}}-r_{o}^{2}\sum_{i=m+1}^{N-1}t_{o}^{j-2(m+1)+i}\hat{X}_{i}^{\mathrm{(in)}}-r_{o}r_{d}t_{o}^{N+j-2(m+1)}\hat{X}_{N}^{\mathrm{(in)}}+\nonumber \\
 & + & r_{o}t_{o}^{j-1}\hat{X}_{A}^{\mathrm{(in)}}+r_{o}t_{d}t_{o}^{N+j-2(m+1)}\hat{X}_{B}^{\mathrm{(in)}}\label{xj2}
\end{eqnarray}
for $m+1\leq j\leq N-1$, and
\begin{eqnarray}
\hat{X}_{N}^{(2)} & = & \left\{ 1-r_{d}^{2}\left[2-t_{o}^{2(N-m-1)}\right]\right\} \hat{X}_{N}^{\mathrm{(in)}}-r_{o}r_{d}\sum_{i=1}^{m}t_{o}^{N-1-i}\hat{X}_{i}^{\mathrm{(in)}}-r_{o}r_{d}\sum_{i=m+1}^{N-1}t_{o}^{N-2(m+1)+i}\hat{X}_{i}^{\mathrm{(in)}}+\nonumber \\
 & - & r_{d}t_{o}^{N-1}\hat{X}_{A}^{\mathrm{(in)}}+r_{d}t_{d}\left[2-t_{o}^{2(N-m-1)}\right]\hat{X}_{B}^{\mathrm{(in)}}\label{xN2}
\end{eqnarray}

At first sight, the target modes are puzzling linear combinations
of all input modes, target and ancillary ones. However, in this stage,
the target modes can be rebuilt using the outcomes of the homodyne
detections to modulate them. In these procedures, the photocurrents
produced by detectors are managed by a circuitry to operate optical
devices, with magneto-optical or electro-optical media, which are
able to modulate the quadrature amplitude components of the laser
beams \citep{Lam97,Yoshikawa08,Shiozawa18,Buchler01,Yoshikawa07,Miyata14,Ukai11,Yokoyama15,Takeda19,Sakaguchi23}. 

In the present scheme, feedforward modulations must be applied on
target modes \eqref{xi2}--\eqref{xN2} by adding the outcomes of
ancillary quadrature measurements, weighted by specific gain, such
that
\begin{equation}
\hat{X}_{j}^{\mathrm{(out)}}=\left(\begin{array}{c}
\hat{q}_{j}^{\mathrm{(out)}}\\
\hat{p}_{j}^{\mathrm{(out)}}
\end{array}\right)=\left(\begin{array}{c}
\hat{q}_{j}^{(2)}+f_{j}\hat{q}_{B}^{\mathrm{(out)}}\\
\hat{p}_{j}^{(2)}+g_{j}\hat{p}_{A}^{\mathrm{(out)}}
\end{array}\right)\label{modulation}
\end{equation}
with the gains $f_{j}$ and $g_{j}$ fixed exactly as previous coefficients
defined in Eqs. \eqref{gi}--\eqref{fN}.

Finally, the produced output target modes are
\begin{equation}
\left(\begin{array}{c}
\hat{q}_{j}^{\mathrm{(out)}}\\
\hat{p}_{j}^{\mathrm{(out)}}
\end{array}\right)=\left(\begin{array}{c}
\hat{q}_{j}^{\mathrm{(in)}}-f_{j}\hat{q}_{A}^{\mathrm{(in)}}\\
\hat{p}_{j}^{\mathrm{(in)}}+2r_{d}t_{d}g_{j}\hat{p}_{N}^{\mathrm{(in)}}+(r_{d}^{2}-t_{d}^{2})g_{j}\hat{p}_{B}^{\mathrm{(in)}}
\end{array}\right)\label{target-out-j}
\end{equation}
 for $1\leq j\leq N-1$, and 
\begin{equation}
\left(\begin{array}{c}
\hat{q}_{N}^{\mathrm{(out)}}\\
\hat{p}_{N}^{\mathrm{(out)}}
\end{array}\right)=\left(\begin{array}{c}
\hat{q}_{N}^{\mathrm{(in)}}-2r_{d}t_{d}\sum_{j=1}^{N-1}g_{j}\hat{q}_{j}^{\mathrm{(in)}}-(r_{d}^{2}-t_{d}^{2})f_{N}\hat{q}_{A}^{\mathrm{(in)}}\\
\hat{p}_{N}^{\mathrm{(in)}}+g_{N}\hat{p}_{B}^{\mathrm{(in)}}
\end{array}\right)\label{target-out-N}
\end{equation}
 for the N-th mode. The ancillary quadrature operators must be considered
components of known off-line squeezed modes, so that $\hat{q}_{A}^{\mathrm{(in)}}=\hat{q}_{A}^{(0)}e^{-s_{A}}$
and $\hat{p}_{B}^{\mathrm{(in)}}=\hat{p}_{B}^{(0)}e^{-s_{B}}$ with
finite squeezing parameters $s_{A}$ and $s_{B}$. In the ideal case,
$s_{A};s_{B}\rightarrow\infty$, so both ancillary terms vanish, resulting
only the target quadrature terms. Comparing Eqs. \eqref{target-out-j}
and \eqref{target-out-N} to the general single-probe multipartite
QND interaction given by Eqs. \eqref{signalmode-1probe} and \eqref{probemode-1probe},
we can make the following correspondence among the coupling parameters
\begin{equation}
G_{j}=-2r_{d}t_{d}g_{j}=\begin{cases}
2r_{o}r_{d}t_{o}^{2m-N-j+1} & ,1\leq j\leq m\\
2r_{o}r_{d}t_{o}^{j-N} & ,m+1\leq j\leq N-1
\end{cases}.\label{qnd-coeficient}
\end{equation}
 Therefore we obtain coupled output target modes according to a multipartite
QND interaction, but with the addition of excess noise provided by
squeezed ancillary quadratures. This is the cost of extracting the
information from the target modes, inevitably the ancillary mode measurements
cause a back-action on them. However we can minimize this measurement
back-action by using squeezed ancillary modes, so that larger squeezing
parameters reduce the disturbance from $\hat{q}_{A}^{\mathrm{(in)}}$
and $\hat{p}_{B}^{\mathrm{(in)}}$. The information extracted from
the target modes carries quantum correlation data between them. In
the next section, we will see that this information deals with the
entanglement among the modes and how we can use it for its certification.

\section{Homodyne detections used for multipartite entanglement certification }

As previously remarked, the ancillary operators $\hat{p}_{A}^{\mathrm{(out)}}$
and $\hat{q}_{B}^{\mathrm{(out)}}$, given by Eqs. \eqref{pain} and
\eqref{qbin}, must be compatible, so that the measurement device
can measure them simultaneously. It is not surprising that such condition
is not satisfied for any values of $t_{o}$ and $t_{d}$, but it is
possible to search for which values are necessary to achieve the compatibility
of the ancillary mode quadratures. Note that both operators $\hat{p}_{A}^{\mathrm{(out)}}$
and $\hat{q}_{B}^{\mathrm{(out)}}$ are linear combinations of input
target and ancillary modes, 
\begin{eqnarray}
\hat{p}_{A}^{\mathrm{(out)}} & = & \mathcal{\hat{A}}_{\mathrm{target}}+\mathcal{\hat{A}}_{\mathrm{ancilla}}\label{comb-pa}\\
\hat{q}_{B}^{\mathrm{(out)}} & = & \mathcal{\hat{B}}_{\mathrm{target}}+\mathcal{\hat{B}}_{\mathrm{ancilla}}\label{comb-qb}
\end{eqnarray}
where $\mathcal{\hat{A}}_{\mathrm{target}}$ and $\mathcal{\hat{B}}_{\mathrm{target}}$
represent the parts with only input target operators and $\mathcal{\hat{A}}_{\mathrm{ancilla}}$
and $\mathcal{\hat{B}}_{\mathrm{ancilla}}$ represent the parts with
only input ancillary operators, respectively. Thus the commutator
between the measured ancillary quadratures splits as 
\begin{equation}
\left[\hat{q}_{B}^{\mathrm{(out)}},\hat{p}_{A}^{\mathrm{(out)}}\right]=\left[\mathcal{\hat{B}}_{\mathrm{target}},\mathcal{\hat{A}}_{\mathrm{target}}\right]+\left[\hat{\mathcal{B}}_{\mathrm{ancilla}},\mathcal{\hat{A}}_{\mathrm{ancilla}}\right].\label{commutator}
\end{equation}
The wanted compatibility is manifested by canceling out the commutation
relation. In fact, it is possible find it with
\begin{equation}
t_{d}=\frac{t_{o}^{m}}{\sqrt{2-t_{o}^{2(N-m-1)}}}\label{compatibility-cond}
\end{equation}
 and also $r_{d}=\pm\sqrt{1-t_{d}^{2}}$, so that the commutators
of the target and ancillary mode parts vanish independently, 
\begin{equation}
\left[\mathcal{\hat{B}}_{\mathrm{target}},\mathcal{\hat{A}}_{\mathrm{target}}\right]=\left[\mathcal{\hat{B}}_{\mathrm{ancilla}},\mathcal{\hat{A}}_{\mathrm{ancilla}}\right]=0.\label{commutator-compatible}
\end{equation}
Thus $\left[\hat{q}_{B}^{\mathrm{(out)}},\hat{p}_{A}^{\mathrm{(out)}}\right]=0$.
Notice that we obtain compatible measurements not only to the output
ancillary modes, but also to the linear combinations of the target
mode quadratures, which can be understood as EPR-like operators. In
turn, the input ancillary quadrature combinations make another pair
of compatible operators and EPR-like too.

From measured ancillary quadratures \eqref{pain} and \eqref{qbin},
it is possible to obtain the respective noise variances in terms of
the input signal modes: 
\begin{equation}
V_{p}^{A}=k_{A}\left[(t_{d}t_{o}^{N-1})^{2}\left\langle \Delta\left(\sum_{j=1}^{N-1}f_{j}\hat{p}_{j}^{\mathrm{(in)}}-f_{N}\hat{p}_{N}^{\mathrm{(in)}}\right)^{2}\right\rangle +\left\langle \Delta\left(t_{d}t_{o}^{N-1}\hat{p}_{A}^{\mathrm{(in)}}+(1-t_{o}^{2m})\hat{p}_{B}^{\mathrm{(in)}}\right)^{2}\right\rangle \right],\label{measurep-in}
\end{equation}
 and 
\begin{equation}
V_{q}^{B}=k_{B}\left[(t_{d}t_{o}^{N-1})^{2}\left\langle \Delta\left(\sum_{j=1}^{N-1}g_{j}\hat{q}_{j}^{\mathrm{(in)}}-g_{N}\hat{q}_{N}^{\mathrm{(in)}}\right)^{2}\right\rangle +\left\langle \Delta\left((1-t_{o}^{2m})\hat{q}_{A}^{\mathrm{(in)}}-t_{d}t_{o}^{N-1}\hat{q}_{B}^{\mathrm{(in)}}\right)^{2}\right\rangle \right].\label{measureq-in}
\end{equation}
 where factors $k_{A}$ and $k_{B}$ are constants determined by the
measuring devices. 

Since the ancillary modes are prepared off-line, the respective variance
parts in equations \eqref{measurep-in} and \eqref{measureq-in} are
previously known. If the input ancillary modes are prepared in a two-mode
squeezed state and adjusted by a suitable phase-sensitive amplification,
the respective variances are as small as larger is the squeezing factor
of the ancillary modes, reducing the equations \eqref{measurep-in}
and \eqref{measureq-in} exclusively to variances of the target modes.
Otherwise, with known independent ancillary modes, the target mode
variances can be inferred. Anyway the target mode variances are accessible
and compatible data.

With the statistical data of linear combinations of the target quadratures,
we can monitor and certify their quantum correlations, before and
after the QND interaction. For N-partite continuous-variable systems,
Teh and Reid have obtained genuine entanglement conditions \citep{Teh14}.
Defining conjugate linear combinations of the multipartite quadrature
operators,
\begin{equation}
\hat{u}=\sum_{j=1}^{N}a_{j}\hat{p}_{j}\label{u-p}
\end{equation}
 and 
\begin{equation}
\hat{v}=\sum_{j=1}^{N}b_{j}\hat{q_{j}},\label{v-q}
\end{equation}
and considering all disjoint bipartitions of $y$ and $z$ modes,
whose union is whole system, so that $y+z=N$, thus a set of numbers
\begin{equation}
S_{B}^{(y,z)}=\biggl|\sum_{k_{y}=1}^{y}a_{k_{y}}b_{k_{y}}\biggr|+\biggl|\sum_{k_{z}=1}^{z}a_{k_{z}}b_{k_{z}}\biggr|,\label{sum-absolute}
\end{equation}
with the index $k_{y}$ running only in bipartition of $y$ modes,
and $k_{z}$ for the bipartition of $z$ modes, can be obtained. It
was proved in reference \citep{Teh14} that the violation of the following
inequality,
\begin{equation}
\langle(\Delta\hat{u})^{2}\rangle+\langle(\Delta\hat{v})^{2}\rangle\geq2\min\left[S_{B}^{(y,z)}\right],\label{non-gme-condition}
\end{equation}
 is a sufficient condition to demonstrate genuine N-partite entanglement.
Still following the mentioned article, it is possible to define the
quantity 
\begin{equation}
\mathrm{Ent}\equiv\frac{\langle(\Delta\hat{u})^{2}\rangle+\langle(\Delta\hat{v})^{2}\rangle}{2\min\left[S_{B}^{(y,z)}\right]},\label{gme-condition}
\end{equation}
 so that the condition $\mathrm{Ent}<1$ certifies the multipartite
genuine entanglement.

We have already showed that, measuring the ancillary mode quadratures
$\hat{p}_{A}^{\mathrm{(out)}}$ and $\hat{q}_{B}^{\mathrm{(out)}}$
, it is possible to obtain the variances of linear combinations of
the target conjugate quadratures \eqref{measurep-in} and \eqref{measureq-in},
in order that they can be applied in expressions \eqref{non-gme-condition}
and \eqref{gme-condition}, identifying the relations
\begin{eqnarray}
\langle(\Delta\hat{u})^{2}\rangle & = & \alpha\left\langle \Delta\left(\sum_{j=1}^{N-1}f_{j}\hat{p}_{j}^{\mathrm{(in)}}-f_{N}\hat{p}_{N}^{\mathrm{(in)}}\right)^{2}\right\rangle ,\label{variance-u}\\
\langle(\Delta\hat{v})^{2}\rangle & = & \beta\left\langle \Delta\left(\sum_{j=1}^{N-1}g_{j}\hat{q}_{j}^{\mathrm{(in)}}-g_{N}\hat{q}_{N}^{\mathrm{(in)}}\right)^{2}\right\rangle ,\label{variance-v}
\end{eqnarray}
 where $\alpha$ and $\beta$ are parameter adjusted by the experimental
apparatus.

Being consequence of the commutation relations \eqref{commutator-compatible},
the constraint $\sum_{j=1}^{N}f_{j}g_{j}=0$ is valid. Substituting
in Eq. \eqref{sum-absolute} , the minimum value for $S_{B}^{(y,z)}$
results in 
\begin{equation}
\min\left[S_{B}^{(y,z)}\right]=2\sqrt{\alpha\beta}\min[|f_{1}g_{1}|,|f_{N-1}g_{N-1}|].\label{min-limit}
\end{equation}
 Thus we have all quantities to calculate the condition \eqref{gme-condition}
with the outcomes obtained by our scheme. So let us study a few particular
cases of calculations of the output target modes and respective genuine
entanglement conditions. In what follows, we will use $\alpha=\beta=1$.

\section{Examples of entanglement certification}

Since we have obtained the output ancillary quadratures \eqref{pain}
and \eqref{qbin}, so that their measurements provide the noise variances
\eqref{measurep-in} and \eqref{measureq-in}, then we can calculate
the variances \eqref{variance-u} and \eqref{variance-v} and the
quantity \eqref{min-limit}. Thus particular cases with different
setups and mode numbers can be studied to obtain a better understanding
of the scheme operation. We shall consider a few genuine multipartite
entangled states to the input modes. It is interesting to use two
well-known states: the CV GHZ and CV EPR-like states \citep{Teh14,Loock00,Armstrong}.
These states can be construct to multipartite systems with three or
more modes. So we will focus on tri and tetrapartite modes and calculate
the parameter ranges needed to obtain violations of condition \eqref{non-gme-condition},
or equivalently obtain $\mathrm{Ent}<1$ in Eq. \eqref{gme-condition},
certifying genuine entanglement. 

\subsection{Tripartite Target Modes}

Following the beam-splitter setup presented in the Section III, which
there are $N-1$ input signal modes and a single probe mode, let us
consider a tripartite input system, so that modes 1 and 2 are the
signal modes and mode 3 is the probe mode. It is also needed to choice
the number $m$ to set up the implementation. For $m=2$, the completion
of the scheme provides the following output signal modes
\begin{equation}
\left(\begin{array}{c}
\hat{q}_{1}^{\mathrm{(out)}}\\
\hat{p}_{1}^{\mathrm{(out)}}
\end{array}\right)=\left(\begin{array}{c}
\hat{q}_{1}^{\mathrm{(in)}}+\frac{r_{o}}{t_{o}}\hat{q}_{A}^{\mathrm{(in)}}\\
\hat{p}_{1}^{\mathrm{(in)}}-2r_{o}r_{d}t_{o}\hat{p}_{3}^{\mathrm{(in)}}-\frac{r_{o}t_{o}}{t_{d}}(r_{d}^{2}-t_{d}^{2})\hat{p}_{B}^{\mathrm{(in)}}
\end{array}\right)\label{example-3-2-x1}
\end{equation}
and
\begin{equation}
\left(\begin{array}{c}
\hat{q}_{2}^{\mathrm{(out)}}\\
\hat{p}_{2}^{\mathrm{(out)}}
\end{array}\right)=\left(\begin{array}{c}
\hat{q}_{2}^{\mathrm{(in)}}+\frac{r_{o}}{t_{o}^{2}}\hat{q}_{A}^{\mathrm{(in)}}\\
\hat{p}_{2}^{\mathrm{(in)}}-2r_{o}r_{d}\hat{p}_{3}^{\mathrm{(in)}}-\frac{r_{o}}{t_{d}}(r_{d}^{2}-t_{d}^{2})\hat{p}_{B}^{\mathrm{(in)}}
\end{array}\right),\label{example-3-2-x2}
\end{equation}
and the output probe mode
\begin{equation}
\left(\begin{array}{c}
\hat{q}_{3}^{\mathrm{(out)}}\\
\hat{p}_{3}^{\mathrm{(out)}}
\end{array}\right)=\left(\begin{array}{c}
\hat{q}_{3}^{\mathrm{(in)}}+2r_{o}r_{d}t_{o}\hat{q}_{1}^{\mathrm{(in)}}+2r_{o}r_{d}\hat{q}_{2}^{\mathrm{(in)}}+\frac{r_{d}}{t_{o}^{2}}(1-2t_{o}^{4})\hat{q}_{A}^{\mathrm{(in)}}\\
\hat{p}_{3}^{\mathrm{(in)}}+\frac{r_{d}}{t_{d}}\hat{p}_{B}^{\mathrm{(in)}}
\end{array}\right).\label{example-3-2-x3}
\end{equation}

As we can notice in Eqs. \eqref{example-3-2-x1}--\eqref{example-3-2-x3},
the output target modes are properly prepared to a further QND measurement
or other similar procedure. The remaining ancillary quadratures $\hat{q}_{A}^{\mathrm{(in)}}$
and $\hat{p}_{B}^{\mathrm{(in)}}$ add noise to resulting system,
but their influence is reduced by using suitable squeezed input ancillary
states.

In turn, the ancillary modes provide the following measured quadratures
\begin{equation}
\hat{q}_{B}^{\mathrm{(out)}}=-r_{o}t_{o}^{2}(t_{o}\hat{q}_{1}^{\mathrm{(in)}}+\hat{q}_{2}^{\mathrm{(in)}})-r_{d}t_{o}^{2}\hat{q}_{3}^{\mathrm{(in)}}-(1-t_{o}^{4})\hat{q}_{A}^{\mathrm{(in)}}+t_{d}t_{o}^{2}\hat{q}_{B}^{\mathrm{(in)}}\label{example-3-2-qb}
\end{equation}
and
\begin{equation}
\hat{p}_{A}^{\mathrm{(out)}}=-r_{o}t_{d}(t_{o}\hat{p}_{1}^{\mathrm{(in)}}+\hat{p}_{2}^{\mathrm{(in)}})+r_{d}t_{d}\hat{p}_{3}^{\mathrm{(in)}}+t_{d}t_{o}^{2}\hat{p}_{A}^{\mathrm{(in)}}+r_{d}^{2}\hat{p}_{B}^{\mathrm{(in)}}.\label{example-3-2-pa}
\end{equation}

To obtain compatible measurement for the quadratures \eqref{example-3-2-qb}
and \eqref{example-3-2-pa}, it is necessary to fix the transmission
coefficients as
\begin{equation}
t_{d}=t_{o}^{2}.\label{example-3-2-comp-cond}
\end{equation}
 Of course, because the ideal relation $t^{2}+r^{2}=1$, the reflection
coefficients are determined too. The Eq. \eqref{min-limit} is reduce
to
\begin{equation}
\min\left[S_{B}^{(y,z)}\right]=2\frac{1-t_{o}^{2}}{t_{o}^{2}}.\label{example-3-2-min-lim}
\end{equation}

\begin{figure}[H]
\begin{centering}
\includegraphics[width=10cm]{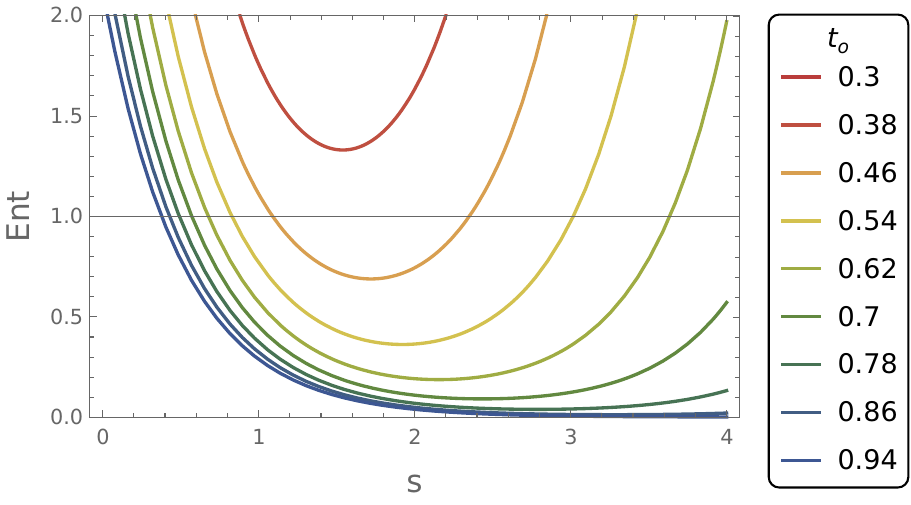}
\par\end{centering}
\centering{}\caption{Certifying genuine tripartite entanglement for input target modes
with CV EPR-type state. If $Ent<1$, then genuine entanglement is
certified. $Ent$ plotted for several transmission coefficient values
$t_{o}$ and in function of the squeezing parameter of the input state,
$s$. \protect\label{fig-tri-epr}}
\end{figure}

\begin{figure}[H]
\begin{centering}
\includegraphics[width=10cm]{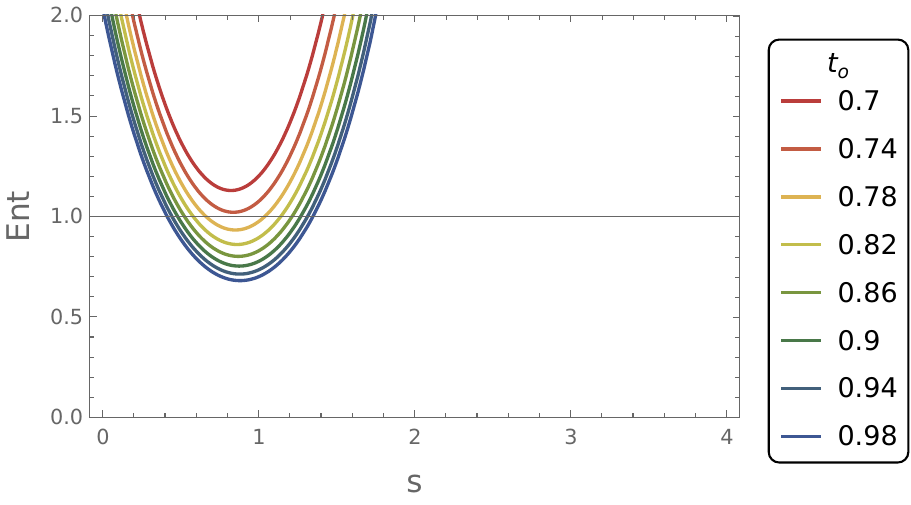}
\par\end{centering}
\centering{}\caption{Certifying genuine tripartite entanglement for input target modes
with CV GHZ state. If $Ent<1$, then genuine entanglement is certified.
$Ent$ plotted for several transmission coefficient values $t_{o}$
and in function of the squeezing parameter of the input state, $s$.
\protect\label{fig-tri-ghz}}
\end{figure}

With the purpose of assessing the scheme effectiveness for different
input states, we have considered two well-known genuine multipartite
entangled systems, the CV EPR-type \citep{Teh14,Armstrong} and the
CV GHZ states \citep{Loock00}. Both cases are widely studied in the
literature. The entanglement certifier $Ent$ can be calculated to
the CV EPR-type and the CV GHZ states and the results are shown in
Figures \ref{fig-tri-epr} and \ref{fig-tri-ghz}, respectively. 

We can notice that the structure and symmetry of the input entangled
states result in different behaviors of the $Ent$ condition. In the
case of tripartite CV EPR-like state, presented in Fig. \ref{fig-tri-epr},
the entanglement certification is obtained to a large range of transmission
coefficient values, $t_{o}$, and the $Ent$ condition advances widely
into the range of values indicating genuine entanglement. On the other
hand, to tripartite CV GHZ state in Fig. \ref{fig-tri-ghz}, the device
setup generates values of $Ent$ much stricter for the genuine entanglement
certification, with a range of transmission coefficient values very
crushed close to 1. 

Therefore, we have seen that the presented scheme executes a tripartite
QND interaction, at the same time it provides measurements for calculating
a genuine entanglement condition. However, the entanglement certification
scheme depends on the transmission coefficient setup of the beam-splitter
operations for different states of the input modes. In fact this feature
is a problem to a static setup, but it can be overcome with dynamic
or adaptive configurations of the beam-splitter operations \citep{Wiseman-Milburn,Miyata16,Hanamura24},
and so to be an advantage, since different input quantum states have
different $Ent$ condition profiles. Thus we can identify different
input entanglement states, plotting their $Ent$ conditions. In Subsection
V.C, we show another scheme setup providing a strong certification
of the multipartite CV GHZ states.

\subsection{Tetrapartite Target Modes}

As shown, the scheme can be applied to multipartite systems of larger
multiplicity. Below we show two cases for tetrapartite input target
modes, which the ancillary modes cross each other between different
beam-splitter couplings with the target modes, featured by parameter
$m$.

\subsubsection{N=4 and m=2}

With tetrapartite target modes, we have $N=4$ and if the ancillary
modes cross each other between target modes 2 and 3, we must have
$m=2$. In this case, the output target modes are
\begin{equation}
\left(\begin{array}{c}
\hat{q}_{j}^{\mathrm{(out)}}\\
\hat{p}_{j}^{\mathrm{(out)}}
\end{array}\right)=\left(\begin{array}{c}
\hat{q}_{j}^{\mathrm{(in)}}-f_{j}\hat{q}_{A}^{\mathrm{(in)}}\\
\hat{p}_{j}^{\mathrm{(in)}}+2r_{d}t_{d}g_{j}\hat{p}_{4}^{\mathrm{(in)}}+(r_{d}^{2}-t_{d}^{2})g_{j}\hat{p}_{B}^{\mathrm{(in)}}
\end{array}\right)\label{example-4-2-xj}
\end{equation}
for $1\leq j\leq3$, and 
\begin{equation}
\left(\begin{array}{c}
\hat{q}_{4}^{\mathrm{(out)}}\\
\hat{p}_{4}^{\mathrm{(out)}}
\end{array}\right)=\left(\begin{array}{c}
\hat{q}_{4}^{\mathrm{(in)}}-2r_{d}t_{d}\sum_{j=1}^{3}g_{j}\hat{q}_{j}^{\mathrm{(in)}}-(f_{4}+2r_{d}t_{o}^{2m-3})\hat{q}_{A}^{\mathrm{(in)}}\\
\hat{p}_{4}^{\mathrm{(in)}}+g_{4}\hat{p}_{B}^{\mathrm{(in)}}
\end{array}\right),\label{example-4-2-x4}
\end{equation}
so that the coupling coefficients are 
\begin{eqnarray}
f_{j} & = & -r_{o}t_{o}^{-j}\label{example-4-2-fj}\\
g_{j} & = & -r_{o}t_{d}^{-1}t_{o}^{-j+1},\label{example-4-2-gj}
\end{eqnarray}
 for $1\leq j\leq2$, 
\begin{eqnarray}
f_{3} & = & -r_{o}t_{o}^{-2}\label{example-4-2-f3}\\
g_{3} & = & -r_{o}t_{d}^{-1}t_{o}^{-1},\label{example-4-2-g3}
\end{eqnarray}
 for $j=3$, and 
\begin{eqnarray}
f_{4} & = & -r_{d}t_{o}^{-3}(2-t_{o}^{2})\label{example-4-2-f4}\\
g_{4} & = & r_{d}t_{d}^{-1},\label{example-4-2-g4}
\end{eqnarray}
for the fourth target mode. As before, the modes from 1 to 3 are the
signal modes and the fourth mode is the probe mode of the tetrapartite
QND coupling. The output ancillary modes provides the following quadratures
measured in the homodyne detection, 
\begin{equation}
\hat{q}_{B}^{\mathrm{(out)}}=-r_{o}t_{o}^{2}(t_{o}\hat{q}_{1}^{\mathrm{(in)}}+\hat{q}_{2}^{\mathrm{(in)}}+\hat{q}_{3}^{\mathrm{(in)}})-r_{d}t_{o}\hat{q}_{4}^{\mathrm{(in)}}-(1-t_{o}^{4})\hat{q}_{A}^{\mathrm{(in)}}+t_{d}t_{o}^{3}\hat{q}_{B}^{\mathrm{(in)}}\label{example-4-2-qb}
\end{equation}
 and 
\begin{equation}
\hat{p}_{A}^{\mathrm{(out)}}=-r_{o}t_{o}t_{d}(t_{o}\hat{p}_{1}^{\mathrm{(in)}}+\hat{p}_{2}^{\mathrm{(in)}}+\hat{p}_{3}^{\mathrm{(in)}})+r_{d}(2-t_{o}^{2})\hat{p}_{4}^{\mathrm{(in)}}+t_{d}t_{o}^{3}\hat{p}_{A}^{\mathrm{(in)}}+[1-t_{d}^{2}(2-t_{o}^{2})]\hat{p}_{B}^{\mathrm{(in)}}.\label{example-4-2-pa}
\end{equation}
 The quadratures \eqref{example-4-2-qb} and \eqref{example-4-2-pa}
commutate each other if the transmission coefficient condition, 
\begin{equation}
t_{d}=\frac{t_{o}^{2}}{\sqrt{2-t_{o}^{2}}},\label{example-4-2-comp-cond}
\end{equation}
 is obeyed.

The tetrapartite genuine entanglement condition \eqref{gme-condition}
is calculated from the target mode terms of quadratures \eqref{example-4-2-qb}
and \eqref{example-4-2-pa}, and obtaining 
\begin{equation}
\min\left[S_{B}^{(y,z)}\right]=\frac{2(1-t_{o}^{2})\sqrt{2-t_{o}^{2}}}{t_{o}^{3}}.\label{example-4-2-min-lim}
\end{equation}

\begin{figure}[H]
\begin{centering}
\includegraphics[width=10cm]{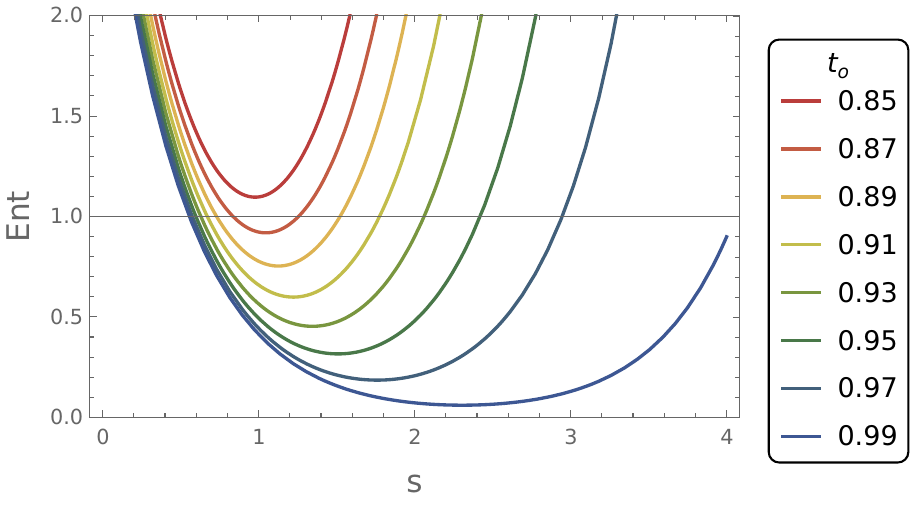}
\par\end{centering}
\centering{}\caption{Certifying genuine tetrapartite entanglement for input target modes
with asymmetric CV EPR-type I state. If $Ent<1$, then genuine entanglement
is certified. $Ent$ plotted for several transmission coefficient
values $t_{o}$ and in function of the squeezing parameter of the
input state, $s$. The scheme used in the plot is configured with
$m=2$, that is, the beam-splitter coupling sequence is symmetric.
\protect\label{fig-tetra2-epr}}
\end{figure}

Thus we can assess the viability of entanglement detection by the
operation of the proposed scheme for the case of tetrapartite modes.
Considering input modes in the CV EPR-type I state, according to the
definition by reference \citep{Teh14}, the $Ent$ condition \eqref{gme-condition}
can be plotted in terms of the transmission coefficient, $t_{o}$,
and the squeezed factor of the input mode state, $s$. The result
is shown in Fig. \ref{fig-tetra2-epr}, where we notice the transmission
coefficient setup for detecting genuine entanglement. There is a limitation
to larger squeezed factors. However, variations of the scheme setup
can improve the detection conditions and overcome this difficulty,
as we can see in the next example. 

\subsubsection{N=4 and m=3}

With the goal of comparing variations of the proposed general scheme,
we consider the tetrapartite case again, but with the ancillary modes
cross each other between target modes 3 and 4, so that it corresponds
to $m=3$. Thus the output target modes are as before in Eqs. \eqref{example-4-2-xj}
and \eqref{example-4-2-x4}, but the coupling coefficients are 
\begin{eqnarray}
f_{j} & = & -r_{o}t_{o}^{-j}\label{example-4-3-fj}\\
g_{j} & = & -r_{o}t_{d}^{-1}t_{o}^{3-j},\label{example-4-3-gj}
\end{eqnarray}
 for $1\leq j\leq3$, and 
\begin{eqnarray}
f_{4} & = & -r_{d}t_{o}^{-3}\label{example-4-3-f4}\\
g_{4} & = & r_{d}t_{d}^{-1},\label{example-4-3-g4}
\end{eqnarray}
for the fourth target mode. The output ancillary quadratures measured
in the homodyne detection are now found to be
\begin{equation}
\hat{q}_{B}^{\mathrm{(out)}}=-r_{o}t_{o}^{3}(t_{o}^{2}\hat{q}_{1}^{\mathrm{(in)}}+t_{o}\hat{q}_{2}^{\mathrm{(in)}}+\hat{q}_{3}^{\mathrm{(in)}})-r_{d}t_{o}^{3}\hat{q}_{4}^{\mathrm{(in)}}-(1-t_{o}^{6})\hat{q}_{A}^{\mathrm{(in)}}+t_{d}t_{o}^{3}\hat{q}_{B}^{\mathrm{(in)}},\label{example-4-3-qb}
\end{equation}
\begin{equation}
\hat{p}_{A}^{\mathrm{(out)}}=-r_{o}t_{d}(t_{o}^{2}\hat{p}_{1}^{\mathrm{(in)}}+t_{o}\hat{p}_{2}^{\mathrm{(in)}}+\hat{p}_{3}^{\mathrm{(in)}})+r_{d}t_{d}\hat{p}_{4}^{\mathrm{(in)}}+t_{d}t_{o}^{3}\hat{p}_{A}^{\mathrm{(in)}}+(1-t_{d}^{2})\hat{p}_{B}^{\mathrm{(in)}}.\label{example-4-3-pa}
\end{equation}

Again, the quadratures \eqref{example-4-3-qb} and \eqref{example-4-3-pa}
must be commutating operators to obtain compatible and simultaneous
measurements. To satisfy this condition we must have 
\begin{equation}
t_{d}=t_{o}^{3}.\label{example-4-3-comp-cond}
\end{equation}

In turn, with the target mode terms of quadratures \eqref{example-4-3-qb}
and \eqref{example-4-3-pa} and calculating 
\begin{equation}
\min\left[S_{B}^{(y,z)}\right]=\frac{2(1-t_{o}^{2})}{t_{o}^{2}},\label{example-4-3-min-lim}
\end{equation}
we obtain the respective tetrapartite genuine entanglement condition
\eqref{gme-condition}. 

\begin{figure}[H]
\begin{centering}
\includegraphics[width=10cm]{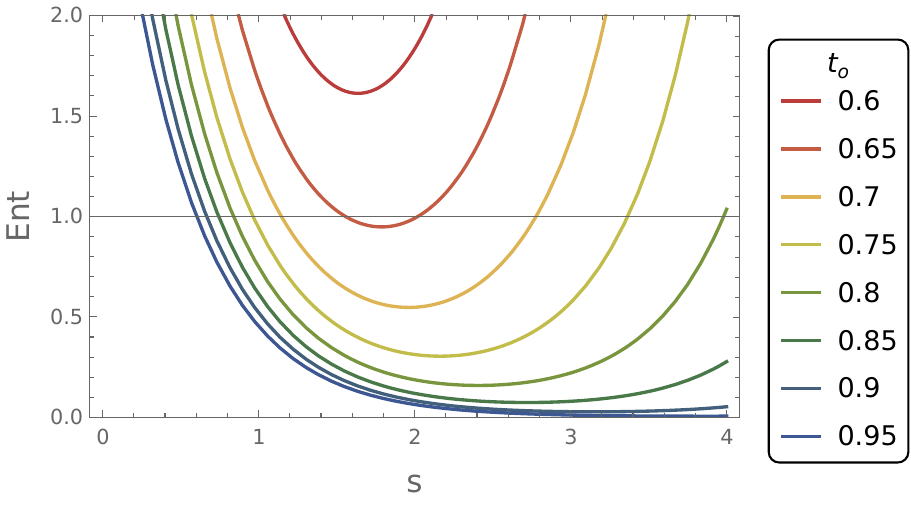}
\par\end{centering}
\centering{}\caption{Certifying genuine tetrapartite entanglement for input target modes
with asymmetric CV EPR-type I state. If $Ent<1$, then genuine entanglement
is certified. $Ent$ plotted for several transmission coefficient
values $t_{o}$ and in function of the squeezing parameter of the
input state, $s$. The scheme used in the plot is configured with
$m=3$, that is, the beam-splitter coupling sequence is asymmetric.
\protect\label{fig-tetra3-epr}}
\end{figure}

Again, considering input modes in the CV EPR-type I state, the $Ent$
condition \eqref{gme-condition} can be plotted in terms of the transmission
coefficient, $t_{o}$, and the squeezed factor of the input mode state,
$s$. We can notice in resulting Fig. \ref{fig-tetra3-epr} that the
entanglement detection is possible for a larger range of parameter
values, $t_{o}$ and $s$, and a more profound values of $Ent$ condition
than the previous setup represented by Fig. \ref{fig-tetra2-epr}.

These examples show the scheme flexibility and possibility of certificating
genuine multipartite entanglement in many different scenarios and
configurations.

\subsection{Other setup to specific states}

As we have seen that the scheme setup defined by conditions \eqref{transmission-o}--\eqref{reflection-d}
entails a poor entanglement certification for input target modes in
the CV GHZ state, that is, the $Ent$ condition has values near to
1 and narrow range of the transmission coefficients. However, other
choices of the transmission coefficients of the beam-splitter operations
can provide different results. Now, let us consider the following
adjustments of the beam-splitter transmission coefficients
\begin{equation}
t_{A,i}=t_{B,j}=t_{B,N}=t_{o}\quad\mbox{, for \ensuremath{i\neq N} and any \ensuremath{j}},\label{transmission-o-alt}
\end{equation}
\begin{equation}
t_{A,N}=t_{d},\label{transmission-d-alt}
\end{equation}
so that the reflection coefficients are $r_{A,i}=r_{B,j}=r_{o}=\sqrt{1-t_{o}^{2}}$,
but beam splitter A,N must have
\begin{equation}
r_{A,N}=-r_{d}=-\sqrt{1-t_{d}^{2}}.\label{reflection-d-alt}
\end{equation}
The conditions \eqref{transmission-o-alt}--\eqref{reflection-d-alt}
differ from previous conditions \eqref{transmission-o}--\eqref{reflection-d}
only in transmission/reflection coefficients of beam splitter B,N.
This single beam-splitter operation choice makes a remarkable difference
in the entanglement detection of specific multipartite states, as
we will see in what follows.

\begin{figure}[H]
\begin{centering}
\includegraphics[width=10cm]{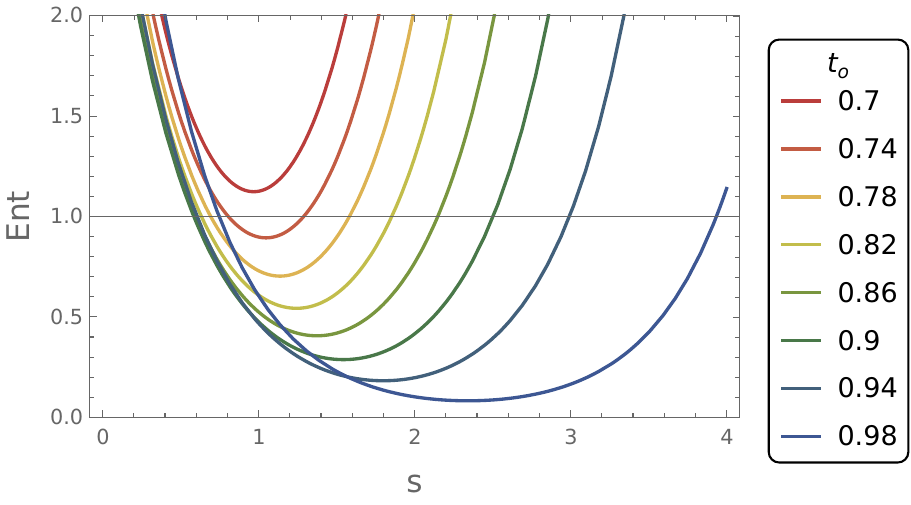}
\par\end{centering}
\centering{}\caption{Certifying genuine tripartite entanglement for input target modes
with CV GHZ state. If $Ent<1$, then genuine entanglement is certified.
$Ent$ plotted for several transmission coefficient values $t_{o}$
and in function of the squeezing parameter of the input state, $s$.
The scheme used in the plot is different from the other Figures, except
Figure 7, by considering only a single beam-splitter operation with
different transmission coefficient $t_{d}$. \protect\label{fig-tri-alt-ghz}}
\end{figure}

The setup defined above generates the output target modes below,
\begin{equation}
\left(\begin{array}{c}
\hat{q}_{1}^{\mathrm{(out)}}\\
\hat{p}_{1}^{\mathrm{(out)}}
\end{array}\right)=\left(\begin{array}{c}
\hat{q}_{1}^{\mathrm{(in)}}+\frac{r_{o}}{t_{o}}\hat{q}_{A}^{\mathrm{(in)}}\\
\hat{p}_{1}^{\mathrm{(in)}}-\frac{r_{o}t_{o}}{t_{d}}(r_{o}t_{d}+t_{o}r_{d})\hat{p}_{3}^{\mathrm{(in)}}-\frac{r_{o}t_{o}}{t_{d}}(r_{o}r_{d}-t_{o}t_{d})\hat{p}_{B}^{\mathrm{(in)}}
\end{array}\right),\label{example-alt-x1}
\end{equation}
\begin{equation}
\left(\begin{array}{c}
\hat{q}_{2}^{\mathrm{(out)}}\\
\hat{p}_{2}^{\mathrm{(out)}}
\end{array}\right)=\left(\begin{array}{c}
\hat{q}_{2}^{\mathrm{(in)}}+\frac{r_{o}}{t_{o}^{2}}\hat{q}_{A}^{\mathrm{(in)}}\\
\hat{p}_{2}^{\mathrm{(in)}}-\frac{r_{o}}{t_{d}}(r_{o}t_{d}+t_{o}r_{d})\hat{p}_{3}^{\mathrm{(in)}}-\frac{r_{o}}{t_{d}}(r_{o}r_{d}-t_{o}t_{d})\hat{p}_{B}^{\mathrm{(in)}}
\end{array}\right),\label{example-alt-x2}
\end{equation}
 and 
\begin{equation}
\left(\begin{array}{c}
\hat{q}_{3}^{\mathrm{(out)}}\\
\hat{p}_{3}^{\mathrm{(out)}}
\end{array}\right)=\left(\begin{array}{c}
\hat{q}_{3}^{\mathrm{(in)}}+\frac{r_{o}}{t_{d}}(r_{o}t_{d}+t_{o}r_{d})\left(t_{o}\hat{q}_{1}^{\mathrm{(in)}}+\hat{q}_{2}^{\mathrm{(in)}}\right)+\left[\frac{r_{o}(1-t_{o}^{4})}{t_{o}^{2}}-\frac{t_{o}^{3}r_{d}}{t_{d}}\right]\hat{q}_{A}^{\mathrm{(in)}}\\
\hat{p}_{3}^{\mathrm{(in)}}+\frac{r_{o}}{t_{o}}\hat{p}_{B}^{\mathrm{(in)}}
\end{array}\right),\label{example-alt-x3}
\end{equation}
 so that, again modes 1 and 2 are the signal modes and mode 3 is the
probe mode. The output ancillary quadratures measured to perform the
process are
\begin{equation}
\hat{q}_{B}^{\mathrm{(out)}}=-r_{o}t_{o}^{2}(t_{o}\hat{q}_{1}^{\mathrm{(in)}}+\hat{q}_{2}^{\mathrm{(in)}}+\hat{q}_{3}^{\mathrm{(in)}})-(1-t_{o}^{4})\hat{q}_{A}^{\mathrm{(in)}}+t_{o}^{3}\hat{q}_{B}^{\mathrm{(in)}}\label{example-alt-qb}
\end{equation}
 and 
\begin{equation}
\hat{p}_{A}^{\mathrm{(out)}}=-r_{o}t_{d}(t_{o}\hat{p}_{1}^{\mathrm{(in)}}+\hat{p}_{2}^{\mathrm{(in)}})+t_{o}r_{d}\hat{p}_{3}^{\mathrm{(in)}}+t_{o}^{2}t_{d}\hat{p}_{A}^{\mathrm{(in)}}+r_{o}r_{d}\hat{p}_{B}^{\mathrm{(in)}}.\label{example-alt-pa}
\end{equation}
Both quadratures are compatible each other if the beam-splitter operation
condition, 
\begin{equation}
t_{d}=\frac{t_{o}}{\sqrt{t_{o}^{2}+(1+t_{o}^{2})(1-t_{o}^{4})}},\label{example-alt-comp-cond}
\end{equation}
 is satisfied. So the genuine entanglement condition \eqref{gme-condition}
is obtained by the target mode terms from Eqs. \eqref{example-alt-qb}
and \eqref{example-alt-pa} and calculating 
\begin{equation}
\min\left[S_{B}^{(y,z)}\right]=\frac{2(1-t_{o}^{2})\sqrt{t_{o}^{2}+(1+t_{o}^{2})(1-t_{o}^{4})}}{t_{o}}.\label{example-alt-min-lim}
\end{equation}
 With the $Ent$ condition, we can plot it for input tripartite modes
in the CV GHZ state. In Fig. \ref{fig-tri-alt-ghz}, we show the $Ent$
plot in terms of the transmission coefficient $t_{o}$ and the squeezed
factor $s$ of the CV GHZ state. 

Comparing with Fig. \ref{fig-tri-ghz} in Subsection V.A., obtained
with the previous scheme setup, the result in Fig. \ref{fig-tri-alt-ghz}
shows that, with simple adjustments, it is possible to improve the
entanglement detection conditions from a restricted situation for
a wide range of parameters $t_{o}$ and $s$, and reaching values
more favorable of the $Ent$ condition, that is, $Ent$ closer to
zero.

\section{Monitoring output signal modes}

We have seen that the proposed scheme can be used as a entanglement
certification procedure. Considering a few genuine entangled states,
namely CV GHZ and EPR-like states, it was shown that the scheme can
detect entanglement with suitable adjusts of the beam-splitter transmission
coefficients. Whereas the detected signals from ancillary modes can
be described in terms of variances of input target modes, providing
data for calculate entanglement conditions, we can also write the
same detected signals in terms of output target modes, just substituting
Eqs. \eqref{target-out-j} and \eqref{target-out-N} in detected ancillary
quadratures \eqref{pain} and \eqref{qbin}. So we obtain
\begin{equation}
\hat{p}_{A}^{\mathrm{(out)}}=t_{d}t_{o}^{N-1}\left[\sum_{j=1}^{N-1}f_{j}\hat{p}_{j}^{\mathrm{(out)}}+(r_{d}^{2}-t_{d}^{2})f_{N}\hat{p}_{N}^{\mathrm{(out)}}\right]+t_{d}t_{o}^{N-1}\hat{p}_{A}^{\mathrm{(in)}}+(1-t_{o}^{2m})\hat{p}_{B}^{\mathrm{(in)}}\label{paout}
\end{equation}
 and 
\begin{equation}
\hat{q}_{B}^{\mathrm{(out)}}=-t_{d}t_{o}^{N-1}\left[(r_{d}^{2}-t_{d}^{2})\sum_{j=1}^{N-1}g_{j}\hat{q}_{j}^{\mathrm{(out)}}+g_{N}\hat{q}_{N}^{\mathrm{(out)}}\right]-\left\{ 1-t_{d}^{2}\left[2-t_{o}^{2(N-m-1)}\right]\right\} \hat{q}_{A}^{\mathrm{(in)}}+t_{d}t_{o}^{N-1}\hat{q}_{B}^{\mathrm{(in)}}.\label{qbout}
\end{equation}
Thus, the detected ancillary quadrature variances in terms of the
output signal modes are written as: 
\begin{equation}
V_{p}^{A}=k_{A}\left[(t_{d}t_{o}^{N-1})^{2}\left\langle \Delta\left(\sum_{j=1}^{N-1}f_{j}\hat{p}_{j}^{\mathrm{(out)}}+(r_{d}^{2}-t_{d}^{2})f_{N}\hat{p}_{N}^{\mathrm{(out)}}\right)^{2}\right\rangle +\left\langle \Delta\left(t_{d}t_{o}^{N-1}\hat{p}_{A}^{\mathrm{(in)}}+(1-t_{o}^{2m})\hat{p}_{B}^{\mathrm{(in)}}\right)^{2}\right\rangle \right],\label{measurep-out}
\end{equation}
and 
\begin{equation}
V_{q}^{B}=k_{B}\left[(t_{d}t_{o}^{N-1})^{2}\left\langle \Delta\left((r_{d}^{2}-t_{d}^{2})\sum_{j=1}^{N-1}g_{j}\hat{q}_{j}^{\mathrm{(out)}}+g_{N}\hat{q}_{N}^{\mathrm{(out)}}\right)^{2}\right\rangle +\left\langle \Delta\left((1-t_{o}^{2m})\hat{q}_{A}^{\mathrm{(in)}}-t_{d}t_{o}^{N-1}\hat{q}_{B}^{\mathrm{(in)}}\right)^{2}\right\rangle \right]\label{measureq-out}
\end{equation}
where $k_{A}$ and $k_{B}$ are constants which depend on detector
and circuitry parameters. As discussed previously, the terms dependent
of the input ancillary modes can be reduced using squeezed modes or
even vanished for specific entangled two-mode squeezed states. Any
way, it is possible to obtain the statistical data from the output
target terms, that is, the variances of operators $\hat{u}$ and $\hat{v}$
in entanglement condition \eqref{gme-condition} are obtained from
\begin{eqnarray}
\hat{u} & = & \sum_{j=1}^{N-1}f_{j}\hat{p}_{j}^{\mathrm{(out)}}+(r_{d}^{2}-t_{d}^{2})f_{N}\hat{p}_{N}^{\mathrm{(out)}},\label{u-p-out}\\
\hat{v} & = & (r_{d}^{2}-t_{d}^{2})\sum_{j=1}^{N-1}g_{j}\hat{q}_{j}^{\mathrm{(out)}}+g_{N}\hat{q}_{N}^{\mathrm{(out)}}.\label{v-q-out}
\end{eqnarray}
 Also, the minimization term $\min\left[S_{B}^{(y,z)}\right]$ is
found as 
\begin{equation}
\min\left[S_{B}^{(y,z)}\right]=2|r_{d}^{2}-t_{d}^{2}|\min[|f_{1}g_{1}|,|f_{N-1}g_{N-1}|].\label{min-limit-out}
\end{equation}

This result is very interesting because we have devised a feasible
scheme for produce a multipartite interaction and simultaneously certificate
the genuine entanglement of the outcome modes. Moreover, the entanglement
certification can be achieved with a computational calculation of
condition \eqref{gme-condition} in time, while the system is on,
that is, the entanglement of the output mode is continuously monitored.
A issue that deserves consideration is the possibility of monitor
the QND interaction between separable signal and probe modes, i. e.,
not entangled. The QND interaction is a entangler unitary operation,
but not produce necessarily genuine entangled estates from input separable
states. For example, taking a single independent probe mode and genuinely
entangled signal modes, the QND interaction does not generally entangle
them, considering the monogamy properties of the genuine entanglement
\citep{Adesso06}. However, given the versatility of use of the QND
interaction, its in-time entanglement certification is a new tool
for the quantum information processing \citep{Friis19}.

Let us consider the first example in Section V, which the scheme is
configured for a tripartite system, with 2 signal modes, 1 probe mode,
and the parameter $m=2$. The gains in the post-modulations are
\begin{eqnarray}
f_{j} & = & -r_{o}t_{o}^{-j},\label{example-out-fj}\\
g_{j} & = & -r_{o}t_{d}^{-1}t_{o}^{2-j},\label{example-out-gj}
\end{eqnarray}
 for $j=1,2$, and 
\begin{eqnarray}
f_{3} & = & -r_{d}t_{o}^{-2},\label{example-out-f3}\\
g_{3} & = & r_{d}t_{d}^{-1}.\label{example-out-g3}
\end{eqnarray}
 The linear combination quadrature operators obtained from the ancillary
mode detections are 
\begin{eqnarray}
\hat{u} & = & -\sqrt{1-t_{o}^{2}}\left[\frac{1}{t_{o}}\hat{p}_{1}^{\mathrm{(out)}}+\frac{1}{t_{o}^{2}}\hat{p}_{2}^{\mathrm{(out)}}+\frac{1-2t_{o}^{4}}{t_{o}^{2}}\hat{p}_{3}^{\mathrm{(out)}}\right],\label{example-out-u}\\
\hat{v} & = & -(1-2t_{o}^{4})\sqrt{1-t_{o}^{2}}\left[\frac{1}{t_{o}}\hat{q}_{1}^{\mathrm{(out)}}+\frac{1}{t_{o}^{2}}\hat{q}_{2}^{\mathrm{(out)}}\right]+\frac{1}{t_{o}^{2}}\hat{q}_{N}^{\mathrm{(out)}},\label{example-out-v}
\end{eqnarray}
 where they are written in function of the output target mode quadratures.
In turn, the minimization of the bipartition function $S_{B}^{(y,z)}$
related to operators \eqref{example-out-u} and \eqref{example-out-v}
is
\begin{equation}
\min\left[S_{B}^{(y,z)}\right]=|1-2t_{o}^{4}|\frac{2(1-t_{o}^{2})}{t_{o}^{2}}.\label{example-out-min-lim}
\end{equation}
 Thus it is possible to establish the entanglement condition \eqref{gme-condition}
for the output target modes.

\begin{figure}[H]
\begin{centering}
\includegraphics[width=10cm]{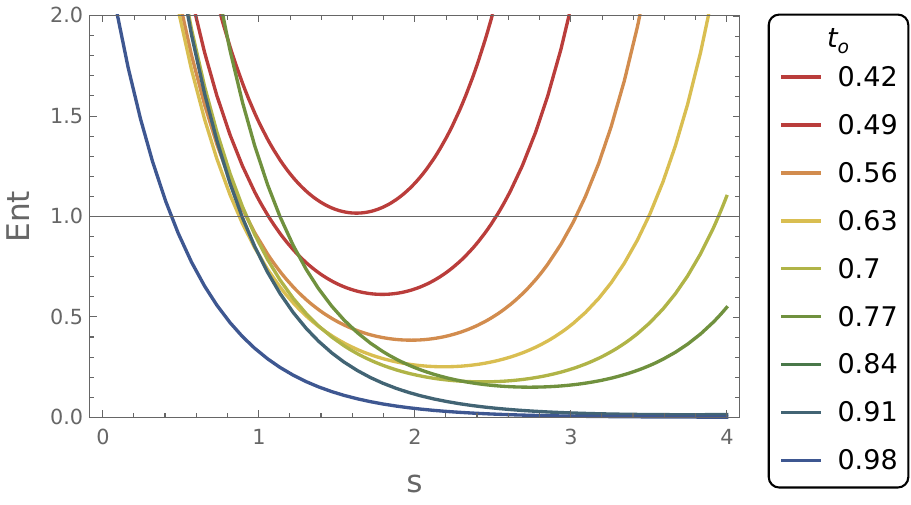}
\par\end{centering}
\centering{}\caption{Certifying genuine tripartite entanglement for output target modes.
The input target modes are in CV EPR-type state. If $Ent<1$, then
genuine entanglement is certified. $Ent$ plotted for several transmission
coefficient values $t_{o}$ and in function of the squeezing parameter
of the input state, $s$. \protect\label{fig-cond-out}}
\end{figure}

Considering as input target modes a tripartite system in CV EPR-like
state, with squeezing factor $s$, it is possible to detect genuine
entanglement adjusting the beam-splitter transmission coefficient
$t_{o}$ in a range of values according to Fig. \ref{fig-cond-out}.

Therefore, we have devised a general scheme that certifies genuine
entanglement of output multipartite modes after a QND coupling among
them. In the presented example, the input modes are in a previous
multipartite entangled state and the QND operation maps the input
modes $\hat{X}_{i}^{\mathrm{(in)}}$, $i=1;2;3$, into Eqs. \eqref{example-3-2-x1}--\eqref{example-3-2-x3}.
Although it is possible to detect genuine entanglement depending on
the scheme parameters, the genuine entangled states of the input and
output mode are different, and it is not obvious what relations exist
between them. The relationship between the input and output entanglement
is a fundamental topic about the multipartite QND coupling, which
will be analyzed in further studies.

\section{Conclusion}

We have presented a general scheme for producing a QND coupling between
two sets of multipartite continuous-variable modes, namely the signal
and probe modes. After developing a general approach to the multipartite
QND interaction, we have focused on the case where the probe set is
composed of a single mode. In this way, it was possible to clearly
show how the output probe mode carries a linear combination of the
observable operators of interest from the signal modes, representing
the information transferred to the probe mode. Of course, the signal
modes carry in their conjugate observable operators the counterpart
of the probe mode, as we expect to happen in the QND process. To perform
the process, the target modes (the signal and probe ones) are coupled
to two ancillary modes, which must be measured by homodyne detection
in conjugate quadratures, producing classical signals used to close
a loop with feedforward modulations. Such ancillary modes introduce
excess noise in the output target modes, but their influence can be
reduced with ancillary modes prepared in suitable squeezed states.
The extension to sets with arbitrary numbers of signal and probe modes
is a natural path. Such studies raise new possibilities and questions
of how the information transferred redundantly to several probe modes
can contribute to further processing, such as metrological precision
improvement \citep{Guo20,Oh20} and fault-tolerant quantum computation
\citep{Takeda19,Sakaguchi23,Menicucci14,Fukui18,Larsen21,Yanagimoto23}. 

We have also seen how the data obtained from the homodyne detections
enable the calculation of conditions for genuine multipartite entanglement,
according to criteria known in the literature \citep{Loock03,He13,Teh14,Teh22,Armstrong}.
In addition to being able to certify the genuine entanglement of the
input target modes, we have noticed that the entanglement detection
profiles depend on the configuration of the beam-splitter operations
and also on the types of states of the input target modes. Although
this may be an additional difficulty for the certification of unknown
generic target modes, on the other hand it opens the possibility of
characterizing the type of entangled states \citep{Friis19,Aolita15}.
In particular, we found that it is possible to distinguish CV GHZ
and EPR-like states to tripartite and tetrapartite states, without
the necessity for complicated state tomography procedures. Finally,
we should highlight the relevance of the capability to monitor genuine
entanglement of the output target modes, after the QND interaction.
This opens the possibility of nondestructive verifications of entanglement
or other correlations during a quantum processing, creating new strategies
in quantum information and computation, which have not yet been explored.


\begin{thebibliography}{99}
\bibitem[(1980)]{Braginsky80} V. B. Braginsky, Yu. I. Vorontsov,
and K. S. Thorne, Science \textbf{209}, 547 (1980).

\bibitem[(1996)]{Braginsky96} V. B. Braginsky and F. Ya. Khalili,
Rev. Mod. Phys. \textbf{68}, 1 (1996).

\bibitem[(1998)]{Grangier98} P. Grangier, J. A. Levenson, and J.-P.
Poizat, Nature \textbf{396}, 537 (1998).

\bibitem[(2010)]{Wiseman-Milburn} H. M. Wiseman and G. J. Milburn,\textit{
Quantum Measurement and Control} (Cambridge University, New York,
2010).

\bibitem[(2018)]{Besse18} J.-C. Besse, S. Gasparinetti, M. C. Collodo,
T. Walter, P. Kurpiers, M. Pechal, C. Eichler, and A. Wallraff, Phys.
Rev. X \textbf{8}, 021003 (2018).

\bibitem[(2020)]{Vasilyev20} D. V. Vasilyev, A. Grankin, M. A. Baranov,
L. M. Sieberer, and P. Zoller, PRX Quantum \textbf{1}, 020302 (2020).

\bibitem[(2020)]{Yoneda20} J. Yoneda, K. Takeda, A. Noiri, T. Nakajima,
S. Li, J. Kamioka, T. Kodera, and S. Tarucha, Nature Comm. \textbf{11},
1144 (2020).

\bibitem[(2007)]{Eckert07} K. Eckert, O. Romero-Isart, M. Rodriguez,
M. Lewenstein, E. S. Polzik, and A. Sanpera, Nature Phys. \textbf{4},
50 (2007).

\bibitem[(2011)]{ReviewMa11} J. Ma, X. Wang, C. P. Sun, and F. Nori,
Phys. Rep. \textbf{509}, 89 (2011). For a review about spin squeezing
by QND measurements, see Section 8.

\bibitem[(2020)]{Yang20} D. Yang, A. Grankin, L. M. Sieberer, D.
V. Vasilyev, and P. Zoller, Nature Comm. \textbf{11}, 775 (2020).

\bibitem[(2005)]{Braunstein05} S. L. Braunstein and P. van Loock,
Rev. Mod. Phys. \textbf{77}, 513 (2005).

\bibitem[(2017)]{Schnabel17} R. Schnabel, Phys. Rep. \textbf{684},
1 (2017).

\bibitem[(2023)]{LIGO23} LIGO O4 Detector Collaboration, Phys. Rev.
X \textbf{13}, 041021 (2023).

\bibitem[(1983)]{Milburn83} G. J. Milburn and D. F. Walls, Phys.
Rev. A \textbf{28}, 2065 (1983).

\bibitem[(1985)]{Imoto85} N. Imoto, H. A. Haus, and Y. Yamamoto,
Phys. Rev. A \textbf{32}, 2287 (1985).

\bibitem[(1986)]{Levenson86} M. D. Levenson, R. M. Shelby, M. Reid,
and D. F. Walls, Phys. Rev. Lett. \textbf{57}, 2473 (1986).

\bibitem[(1985)]{Yurke85} B. Yurke, J. Opt. Soc. Am. B \textbf{2},
732 (1985).

\bibitem[(1989)]{La=000020Porta89} A. La Porta, R. E. Slusher, and
B. Yurke, Phys. Rev. Lett. \textbf{62}, 28 (1989).

\bibitem[(1997)]{Scully} M. O. Scully and M. S. Zubairy, \textit{Quantum
Optics} (Cambridge University, 1997).

\bibitem[(1997)]{Lam97} P. K. Lam, T. C. Ralph, E. H. Huntington,
and H.-A. Bachor, Phys. Rev. Lett. \textbf{79}, 1471 (1997).

\bibitem{Ralph97} T. C. Ralph, Phys. Rev. A \textbf{56}, 4187 (1997).

\bibitem[(2002)]{Andersen02} U. L. Andersen, B. C. Buchler, H.-A.
Bachor, and P. K. Lam, J. Opt. B: Quantum Semiclass. Opt. \textbf{4},
S229 (2002).

\bibitem[(2005)]{Filip05} R. Filip, P. Marek, and U. L. Andersen,
Phys. Rev. A \textbf{71}, 042308 (2005).

\bibitem[(2008)]{Yoshikawa08} J. Yoshikawa, Y. Miwa, A. Huck, U.
L. Andersen, P. van Loock, and A. Furusawa, Phys. Rev Lett. \textbf{101},
250501 (2008).

\bibitem[(2010)]{Marek10} P. Marek and R. Filip, Phys. Rev. A \textbf{81},
042325 (2010).

\bibitem[(2018)]{Shiozawa18} Y. Shiozawa, J. Yoshikawa, S. Yokoyama,
T. Kaji, K. Makino T. Serikawa, R. Nakamura, S. Suzuki, S. Yamazaki,
W. Asavanant, S. Takeda, P. van Loock, and A. Furusawa, Phys. Rev.
A \textbf{98}, 052311 (2018).

\bibitem[(2001)]{Buchler01} B. C. Buchler, P. K. Lam, H.-A. Bachor,
U. L. Andersen, and T. C. Ralph, Phys. Rev. A \textbf{65}, 011803R
(2001).

\bibitem[(2007)]{Yoshikawa07} J. Yoshikawa, T. Hayashi, T. Akiyama,
N. Takei, A. Huck, U. L. Andersen, and A. Furusawa, Phys. Rev. A \textbf{76},
060301(R) (2007).

\bibitem[(2014)]{Miyata14} K. Miyata, H. Ogawa, P. Marek, R. Filip,
H. Yonezawa, J. Yoshikawa, and A. Furusawa, Phys. Rev. A \textbf{90},
060302(R) (2014).

\bibitem[(2011)]{Ukai11} R. Ukai, N. Iwata, Y. Shimokawa, S. C. Armstrong,
A. Politi, J. Yoshikawa, P. van Loock, and A. Furusawa, Phys. Rev.
Lett. \textbf{106}, 240504 (2011).

\bibitem[(2015)]{Yokoyama15} S. Yokoyama, R. Ukai, S. C. Armstrong,
J. Yoshikawa, P. van Loock, and A. Furusawa, Phys. Rev. A \textbf{92},
032304 (2015).

\bibitem[(2019)]{Takeda19} S. Takeda and A. Furusawa, APL Photon.
\textbf{4}, 060902 (2019).

\bibitem[(2023)]{Sakaguchi23} A. Sakaguchi, S. Konno, F. Hanamura,
W. Asavanant, K. Takase, H. Ogawa, P. Marek, R. Filip, J. Yoshikawa,
E. Huntington, H. Yonezawa, and A. Furusawa, Nature Comm. \textbf{14},
3817 (2023).

\bibitem[(2011)]{Marek11} P. Marek, R. Filip, and A. Furusawa, Phys.
Rev. A \textbf{84}, 053802 (2011).

\bibitem[(2016)]{Miyata16} K. Miyata, H. Ogawa, P. Marek, R. Filip,
H. Yonezawa, J. Yoshikawa, and A. Furusawa, Phys. Rev. A \textbf{93},
022301 (2016).

\bibitem[(2018)]{Marek18} P. Marek, R. Filip, H. Ogawa, A. Sakaguchi,
S. Takeda, J. Yoshikawa, and A. Furusawa, Phys. Rev. A \textbf{97},
022329 (2018).

\bibitem[(2019)]{Sefi19} S. Sefi, P. Marek and R. Filip, New J. Phys.
\textbf{21}, 063018 (2019).

\bibitem[(2016)]{Faria16} A. J. de Faria, Phys. Rev. A \textbf{94},
012301 (2016).

\bibitem[(2009)]{Horodecki09} R. Horodecki, P. Horodecki, M. Horodecki,
and K. Horodecki, Rev. Mod. Phys. \textbf{81}, 865 (2009).

\bibitem[(2020)]{Uola20} R. Uola, A. C. S. Costa, H. C. Nguyen, and
O. Gühne, Rev. Mod. Phys. \textbf{92}, 015001 (2020).

\bibitem[(2001)]{Giedke01} G. Giedke, B. Kraus, M. Lewenstein, and
J. I. Cirac, Phys. Rev. A \textbf{64}, 052303 (2001).

\bibitem[(2003)]{Loock03} P. van Loock and A. Furusawa, Phys. Rev.
A \textbf{67}, 052315 (2003).

\bibitem[(2013)]{Sperling13} J. Sperling and W. Vogel, Phys. Rev.
Lett. \textbf{111}, 110503 (2013).

\bibitem[(2013)]{He13} Q. Y. He and M. D. Reid, Phys. Rev. Lett \textbf{111},
250403 (2013).

\bibitem[(2014)]{Teh14} R. Y. Teh and M. D. Reid, Phys. Rev. A \textbf{90},
062337 (2014).

\bibitem[(2015)]{Shchukin15} E. Shchukin and P. van Loock, Phys.
Rev. A \textbf{92}, 042328 (2015).

\bibitem[(2022)]{Teh22} R. Y. Teh, M. Gessner, M. D. Reid, and M.
Fadel, Phys. Rev. A \textbf{105}, 012202 (2022).

\bibitem[(2000)]{Simon00} R. Simon, Phys. Rev. Lett. \textbf{84},
2726 (2000).

\bibitem[(2000)]{DGCZ00} L.-M. Duan, G. Giedke, J. I. Cirac, and
P. Zoller, Phys. Rev. Lett. \textbf{84}, 2722 (2000).

\bibitem[(2003)]{Giovannetti03} V. Giovannetti, S. Mancini, D. Vitali,
and P. Tombesi, Phys. Rev. A \textbf{67}, 022320 (2003).

\bibitem[(2024)]{Hanamura24} F. Hanamura, W. Asavanant, H. Nagayoshi,
A. Sakaguchi, R. Ide, K. Fukui, P. van Loock, and A. Furusawa, Phys.
Rev. A \textbf{110}, 022614 (2024).

\bibitem[(2015)]{Armstrong} S. Armstrong, M. Wang, R. Y. Teh, Q.
Gong, Q. He, J. Janousek, H.-A. Bachor, M. D. Reid, and P. K. Lam,
Nature Phys. \textbf{11}, 167 (2015).

\bibitem[(2000)]{Loock00} P. van Loock and S. L. Braunstein, Phys.
Rev. Lett. \textbf{84}, 3482 (2000).

\bibitem[(2006)]{Adesso06} G. Adesso and F. Illuminati, New J. Phys.
\textbf{8}, 15 (2006).

\bibitem[(2019)]{Friis19} N. Friis, G. Vitagliano, M. Malik, and
M. Huber, Nature Rev. Phys. \textbf{1}, 72 (2019).

\bibitem[(2011)]{Giovannetti11} V. Giovannetti, S. Lloyd, and L.
Maccone, Nature Photon. \textbf{5}, 222 (2011)

\bibitem[(2020)]{Guo20} X. Guo, C. R. Breum, J. Borregaard, S. Izumi,
M. V. Larsen, T. Gehring, M. Christandl, J. S. Neergaard-Nielsen,
and U. L. Andersen, Nature Phys. \textbf{16}, 281 (2020).

\bibitem[(2020)]{Oh20} C. Oh, C. Lee, S. H. Lie, and H. Jeong, Phys.
Rev. Res. \textbf{2}, 023030 (2020).

\bibitem[(2014)]{Menicucci14} N. C. Menicucci, Phys. Rev. Lett. \textbf{112},
120504 (2014).

\bibitem[(2018)]{Fukui18} K. Fukui, A. Tomita, A. Okamoto, and K.
Fujii, Phys. Rev. X \textbf{8}, 021054 (2018).

\bibitem[(2021)]{Larsen21} M. V. Larsen, X. Guo , C. R. Breum, J.
S. Neergaard-Nielsen, and U. L. Andersen, Nature Phys. \textbf{17},
1018 (2021).

\bibitem[(2023)]{Yanagimoto23} R. Yanagimoto, R. Nehra, R. Hamerly,
E. Ng, A. Marandi, and H. Mabuchi, Phys. Rev. X Quantum \textbf{4},
010333 (2023).

\bibitem[(2015)]{Aolita15} L. Aolita, C. Gogolin, M. Kliesch, and
J. Eisert, Nature Comm. \textbf{6}, 8498 (2015).

\end{thebibliography}
\end{document}